\documentclass[a4paper,pra,aps,12pt,nofootinbib]{revtex4}

\usepackage[T1]{fontenc}
\usepackage[utf8]{inputenc}
\usepackage{units}
\usepackage{textcomp}
\usepackage{amsmath}    
\usepackage{graphicx}   
\usepackage{verbatim}   
\usepackage{color}      
\usepackage{subfigure}  
\usepackage{hyperref}   
\hypersetup{citebordercolor=.2 .8 .2,urlbordercolor=0 .75 1}
\usepackage{nicefrac}
\usepackage{amsfonts}
\usepackage{amsthm}
\usepackage{amssymb}
\usepackage{booktabs}
\usepackage{braket}
\usepackage{fancyhdr}
\usepackage{moreverb}
\usepackage[brazil]{babel}
\usepackage{mathtools}
\usepackage{slashed}

\usepackage{pgfplots}
\pgfplotsset{compat=newest}

\makeatletter
\renewcommand*{\email}[1][E-mail: ]{\begingroup\sanitize@url\@email{#1}}
\makeatother

\begin{document}

	\title{Quantum aspects of the classical Maxwell's equations in free space from the perspective of the generalised correspondence principle}

	\author{M. F. Araujo de Resende}
	\email{resende@if.usp.br}
	\affiliation{Instituto de Física, Universidade de São Paulo, 05508-090 São Paulo SP, Brasil}
	\affiliation{Centro de Ciências Naturais e Humanas, Universidade Federal do ABC, 09210-580 Santo André SP, Brasil}
	
	\author{Leonardo S. F. Santos}
	\email{leonardo.sioufi@unifesp.br}
	\affiliation{Departamento de Física, Universidade Federal de São Paulo, 09920-000 Diadema SP, Brasil}
	
	\author{R. Albertini Silva}
	\email{ralberti@if.usp.br}
	\affiliation{Instituto de Física, Universidade de São Paulo, 05508-090 São Paulo SP, Brasil}

	\date{\today }

	\begin{abstract}
		Due to the advent of Quantum Mechanics' 100th anniversary in 2025, we wrote this review paper in order to present a discussion that addresses the foundations of this theory. And since the creation of this Mechanics and other quantum theories was guided, for instance, by correspondence principles that needed to be identified between them and other well-established physical theories, this paper will be devoted to discussing the correspondence between Quantum Mechanics and Maxwell's theory of electromagnetism. More precisely, what we will do throughout this paper is discuss how the Maxwell's electromagnetic theory in free space and the generalised formulation of the correspondence principle already pointed, together, to the basis of a Quantum Mechanics that would only be formulated half a century later. And, in order to make this very clear, we will show that the quantum-mechanical description of a photon can already be identified, simply, by manipulating Maxwell's equations in free space with mathematical resources that, for instance, were already well known before the advent of Quantum Mechanics.
	\end{abstract}

	\maketitle

	\section{\label{sec01}Introduction}
	
		If, for some reason, the scientific community needed to choose the greatest climaxes in the history of Physics, most of them would certainly date from the beginning of the 19th Century until the first half of the 20th Century. After all, it was precisely in this period that some theories, which have a crucial importance in Physics, emerged and broke some paradigms. And one of these theories was the dynamical theory of the electromagnetic field developed by James Clerk Maxwell (1831 -- 1879) that, among other things, demonstrated to everyone that light could actually be interpreted as an electromagnetic wave \cite{maxwell-wave}.
		
		By the way, the brilliance and adequacy of the electromagnetic theory developed by Maxwell were so great that it not only unleashed a great technological revolution \cite{shaver}, but also to give rise to new theories shortly thereafter. And one of these new theories, which emerged in the same century, was proposed by Hendrik Lorentz (1853 -- 1928), who tried to describe the propagation of light through a luminiferous aether \cite{lorentz-1,lorentz-2,lorentz-3,lorentz-4,lorentz-5,lorentz-6,lorentz-7,lorentz-8}. Incidentally, Maxwell himself was also believed in the existence of this aether since, for instance, all the non-luminous waves known at that time needed a material medium to propagate \cite{whittaker,theocharis,dolling}. However, in spite of the mathematical foundation of Lorentz's theory being flawless, the only (and big!) problem that did not allow it to be interpreted as a physical theory was precisely the impossibility of detecting (experimentally) the existence of this aether \cite{fizeau-1,fizeau-2,michelson-morley,michelson-et-al,shankland}. 
		
		Given this impossibility of detecting such aether, the entire mathematical foundation of Lorentz's theory was adapted to a new (and more practical) proposal that Albert Einstein (1879 -- 1955) made in 1905 \cite{einstein-relativity,bohm}. In this new proposal, Einstein abolished the idea of this aether in order to base the mathematical foundation of Lorentz's theory only on two physical principles: (i) the laws of physics should be independent of the uniform motion of an inertial frame of reference; and (ii) the speed of light should have the same constant value in any such frame \cite{penrose}. The implications of this new Lorentz-Einstein theory \cite{miller} (which became popularly known as special theory of relativity) were extensively investigated and, due to the agreement of its predictions with the various theoretical and experimental tests, it not only was accepted by the scientific community, but also ended up being used as a first step towards a more general theory of relativity, where gravitational effects need to be considered \cite{einstein-gr-1,einstein-gr-2,einstein-gr-3}.
		
		However, although classical electromagnetism and special relativity have revealed properties of light that look perfect and non-conflicting, Einstein also published, in that same year of 1905, another paper that blew the minds of those who were already accustomed to the idea that light was a wave \cite{einstein-photon}. After all, by reviewing a paper that Max Planck (1858 -- 1947) had published, in 1901, in order to explain the black body radiation \cite{planck}, Einstein came to a curious conclusion: while Planck assumed (even without knowing how to explain why) that the walls of a black body emitted, when heated, a spectrum of electromagnetic radiation whose energies
		\begin{equation}
			E = hf \label{energia}
		\end{equation}
		were proportional to the frequencies ($ f $) of the electromagnetic waves that describe it, by the same constant ($ h $), Einstein concluded that it was electromagnetic radiation itself that has this quantized nature. By according to this curious conclusion, not only monochromatic lights, but also all other electromagnetic waves that describe non-visible radiation (whose existence was also inferred by Maxwell) could be interpreted as a swarm of discrete energy packets (i.e., particles that have this energy (\ref{energia})) since it was exactly this interpretation that could explain, for instance, the photoelectric effect. And as this Einstein's conclusion was so successful that it won him the Nobel Prize in Physics, in 1921, precisely for his discovery of the law of the photoelectric effect, one of the new question that needed to be answered by the scientific community was: how to conciliate the interpretation of any radiation that can be interpreted in terms of an electromagnetic wave with this new Einstein's proposal, where this same radiation needs to be described as a stream of particles? 
		
		In order to understand the answer to this question, it is important to note that, since the second half of the 19th Century, several researches have been done in pursuit of some new theory that could explain some microscopic phenomena \cite{beer,angstrom-1,kirchhoff,kirch-bunsen-1,kirch-bunsen-2,angstrom-2,balmer,lockyer,rydberg,thomson}. And as these early researches sought to describe the atomic and molecular behaviour of various materials, it is not wrong to say that all these early researches did not deal with relativistic considerations because, for instance, atoms are massive. Nevertheless, by showing that any electromagnetic radiation could be interpreted in this quantum way, Einstein ended up making it quite clear that everything, which was already being developed to describe the microscopic phenomena, also needed to be mixed, in some way, not only with this corpuscular interpretation of electromagnetic radiation, but also with all the content of the electromagnetic theory and special relativity because, among other things,
		\begin{itemize}
			\item it was already known that an atom could be unequivocally characterized by its light spectrum \cite{krishnan}, and
			\item all the particles that compose the atoms/molecules (i.e., protons, neutrons and electrons) interact with the electromagnetic fields that are described by Maxwell's equations \cite{ashcroft}.
		\end{itemize}
		
		Given this new scenario, it was pretty clear that any theory, which was able to describe electromagnetic radiation in this quantum way revealed by Einstein, needed not only to respect the Maxwell's formulation of electromagnetism, but also to be necessarily relativistic from its inception. However, as unexpected as it was for the scientific community to revive the wave-particle debate on the nature of light, the fact is that, in addition to Maxwell's equations already agreed with special relativity \cite{schwarz}, it is possible to prove that these equations also already agreed with what was yet to be formulated by Werner Heisenberg (1901 -- 1976), Erwin Schr\"{o}dinger (1887 -- 1961) and the other physicists more than half a century later. In particular, it is possible to show that Maxwell's electromagnetic theory already supports some quantum description of light and other electromagnetic radiations since, among other things, part of Maxwell's equations in free space takes the same form as the Schr\"{o}dinger equation \cite{schrodinger-1,schrodinger-2}. In this way, as this fact is a little overshadowed in the most references that introduce Quantum Electrodynamics (QED) and, in most cases, it is not even mentioned in Quantum Mechanics textbooks, we can affirm that the basic purpose of this paper is to present a critical and pedagogical review on this subject in order to clarify, for instance, how the Maxwell's electromagnetic theory in free space and the generalised correspondence principle \cite{post,bokulich-1} already pointed, together, to the basis of a Quantum Mechanics that was only be formulated half a century later. 
    
    \section{\label{sec02}Maxwell's equations written in $ \boldsymbol{k} $-space} 
    
    		The first thing we need to do here, in order to begin to understand why Maxwell's equations in free space already support some quantum description, is to recognize, for example, that all the electromagnetic waves that actually exist in Nature are not really flat and monochromatic. By the way, note that some of the general simultaneous solutions of these Maxwell's equations can be written as\footnote{Although we are not concerned with writing the domain of integration in (\ref{relation-rk}), it is worth observing that such a domain is the entire $ \boldsymbol{k} $-space. A similar comment will also apply to other integrals that are written without detailing their domains of integration.} \cite{bracewell}
		\begin{equation}
			\boldsymbol{E} \left( \boldsymbol{r} , t \right) = \int \boldsymbol{\mathcal{E}} _{0} \left( \boldsymbol{k} \right) e^{i \left[ \boldsymbol{k} \cdot \boldsymbol{r} - \omega \left( \boldsymbol{k} \right) t \right] } \ d \boldsymbol{k} \quad \textnormal{and} \quad \boldsymbol{B} \left( \boldsymbol{r} , t \right) = \int \boldsymbol{\mathcal{B}} _{0} \left( \boldsymbol{k} \right) e^{i \left[ \boldsymbol{k} \cdot \boldsymbol{r} - \omega \left( \boldsymbol{k} \right) t \right] } \ d \boldsymbol{k} ,  \label{relation-rk}
		\end{equation}
		among which are the so-called \emph{wave packets}: i.e., the electromagnetic fields $ \left( \boldsymbol{E} \left( \boldsymbol{r} , t \right) , \boldsymbol{B} \left( \boldsymbol{r} , t \right) \right) $ that can be modelled by the same superpositions (\ref{relation-rk}), but where the functions $ \boldsymbol{\mathcal{E}} _{0} \left( \boldsymbol{k} \right) $ and $ \boldsymbol{\mathcal{B}} _{0} \left( \boldsymbol{k} \right) $ are proportional to
		\begin{equation*}
			\exp \left[ - \frac{\left\vert \boldsymbol{k} - \boldsymbol{k} _{0} \right\vert ^{2}}{2 \left( \Delta k \right) ^{2}} \right] \ .
		\end{equation*}
		In other words, $ \boldsymbol{\mathcal{E}} _{0} \left( \boldsymbol{k} \right) $ and $ \boldsymbol{\mathcal{B}} _{0} \left( \boldsymbol{k} \right) $ are two \emph{Gaussian functions} \cite{abramowitz} in this wave packet case, which have the same maximum point $ \boldsymbol{k} = \boldsymbol{k_{0}} $, where $ \Delta k $ represents their peak width.
		
		Regarding the function $ \omega \left( \boldsymbol{k} \right) $ that appears in (\ref{relation-rk}), it is correct to say that, in addition to it being interpreted as the angular frequency of the components
		\begin{equation}
			\boldsymbol{\mathcal{E}} \left( \boldsymbol{k} , t \right) = \boldsymbol{\mathcal{E}} _{0} \left( \boldsymbol{k} \right) e^{- i \omega \left( \boldsymbol{k} \right) \cdot t} \quad \textnormal{and} \quad \boldsymbol{\mathcal{B}} \left( \boldsymbol{k} , t \right) = \boldsymbol{\mathcal{B}} _{0} \left( \boldsymbol{k} \right) e^{- i \omega \left( \boldsymbol{k} \right) \cdot t} \ , \label{eb-identification}
		\end{equation}
		it can also be interpreted as the function that models the \emph{dispersion} of an electromagnetic wave packet. And as this last interpretation allows us to recognize that the group velocity of this wave packet is
		\begin{equation*}
			\boldsymbol{v} _{g} = \boldsymbol{\nabla } _{\boldsymbol{k}} \ \omega \left( \boldsymbol{k} \right) \ ,
		\end{equation*}
		this allows us to conclude, for example, that such dispersion can be approximated by
		\begin{equation*}
			\omega \left( \boldsymbol{k} \right) \ \approx \ \omega _{0} + \left( \boldsymbol{k} - \boldsymbol{k} _{0} \right) ^{T} \cdot \boldsymbol{v} _{g} + \frac{1}{2} \left( \boldsymbol{k} - \boldsymbol{k} _{0} \right) ^{T} \cdot \boldsymbol{\mathsf{H}} _{\omega } \cdot \left( \boldsymbol{k} - \boldsymbol{k} _{0} \right) \ , 
		\end{equation*}
		where $ \boldsymbol{\mathsf{H}} _{\omega } $ is the Hessian of $ \omega \left( \boldsymbol{k} \right) $ \cite{matrix-lewis}. Therefore, by noting that, under a change of variable $ \boldsymbol{k} \ \rightarrow \ \boldsymbol{\kappa } = \boldsymbol{k} - \boldsymbol{k} _{0} $, the expressions of these wave packets can be rewritten as
		\begin{eqnarray*}
			\boldsymbol{E} \left( \boldsymbol{r} , t \right) & = & e^{i \boldsymbol{k} _{0} \boldsymbol{\cdot r}} \int \boldsymbol{\mathcal{E}} _{0} \left( \boldsymbol{k} _{0} + \boldsymbol{\kappa } \right) e^{i \left[ \boldsymbol{\kappa } \cdot \boldsymbol{r} - \omega \left( \boldsymbol{\kappa } \right) t \right] } \ d \boldsymbol{\kappa } \quad \textnormal{and} \\
			\boldsymbol{B} \left( \boldsymbol{r} , t \right) & = & e^{i \boldsymbol{k} _{0} \boldsymbol{\cdot r}} \int \boldsymbol{\mathcal{B}} _{0} \left( \boldsymbol{k} _{0} + \boldsymbol{\kappa } \right) e^{i \left[ \boldsymbol{\kappa } \cdot \boldsymbol{r} - \omega \left( \boldsymbol{\kappa } \right) t \right] } \ d \boldsymbol{\kappa } \ ,
		\end{eqnarray*}
		we can conclude that they are all are, for instance, proportional to
		\begin{equation*}
			e^{i \left( \boldsymbol{k} _{0} \cdot \boldsymbol{r} - \omega _{0} t \right) } \int e^{i \boldsymbol{\kappa } ^{T} \cdot \left( \boldsymbol{r} - \boldsymbol{r} _{0} - \boldsymbol{v} _{g} t \right) } \exp \left[ \frac{\boldsymbol{\kappa } ^{T}}{2} \cdot \left( \frac{1}{\left( \Delta k \right) ^{2}} + i \ \boldsymbol{\mathsf{H}} _{\omega } t \right) \cdot \boldsymbol{\kappa } \right] \ d \boldsymbol{\kappa } \ .
		\end{equation*}
		In this fashion, since this last result makes clear that the modules of this electromagnetic are proportional to
		\begin{equation*}
			\exp \left[ - \frac{\left\vert \boldsymbol{r} - \boldsymbol{r} _{0} - \boldsymbol{v} _{g} t \right\vert ^{2}}{2 \left( \Delta r \right) ^{2}} \right] \ ,
		\end{equation*}
		it also becomes clear that 
		\begin{equation}
			\Delta r = \frac{1}{2 \ \Delta k} \ \sqrt{1 + \left( \boldsymbol{\kappa } ^{T} \cdot \boldsymbol{\mathsf{H}} _{\omega } \cdot \boldsymbol{\kappa } \right) ^{2} t^{2}} \quad \Leftrightarrow \quad \Delta r \cdot \Delta k = \frac{1}{2} \ \sqrt{1 + \left( \boldsymbol{\kappa } ^{T} \cdot \boldsymbol{\mathsf{H}} _{\omega } \cdot \boldsymbol{\kappa } \right) ^{2} t^{2}} \geqslant \frac{1}{2} \ . \label{incerteza}
		\end{equation} 
		
		In view of this last result (\ref{incerteza}), it is quite tempting to conclude, for instance, that it has some relationship with the Heisenberg's uncertainty principle \cite{heisenberg-1}
		\begin{equation}
			\Delta \boldsymbol{r} \cdot \Delta \boldsymbol{p} \geqslant \frac{\hbar }{2} \label{incerteza-heisenberg}
		\end{equation}
		in the electromagnetic context. After all, if this relationship actually exists in order to, for example, allows us to recognize (\ref{incerteza}) and (\ref{incerteza-heisenberg}) as equivalent, this could be of utmost importance in view of the proposal of this paper. However, as all the algebraic manipulations we have done so far still do not endorse this equivalence in the electromagnetic context, it is essential to evaluate whether the replacement of these expressions in Maxwell's equations\footnote{Here, the variable $ \boldsymbol{r} $ that indexes the operator $ \boldsymbol{\nabla } $ represents the differentiation variable. An analogous comment applies when this same operator is indexed by other variables.} \cite{jackson}
    		\begin{subequations} \label{maxwell}
			\begin{align}
				\boldsymbol{\nabla } _{\boldsymbol{r}} \cdot \boldsymbol{E} & = \frac{\rho }{\varepsilon _{0}} \label{maxwell-1} \ , \\
				\boldsymbol{\nabla } _{\boldsymbol{r}} \cdot \boldsymbol{B} & = 0 \label{maxwell-2} \ , \\
				\boldsymbol{\nabla } _{\boldsymbol{r}} \times \boldsymbol{E} & = - \frac{\partial \boldsymbol{B}}{\partial t} \label{maxwell-3} \quad \textnormal{and} \\
				\boldsymbol{\nabla } _{\boldsymbol{r}} \times \boldsymbol{B} & = \mu _{0} \boldsymbol{J} + \frac{1}{c^{2}} \frac{\partial \boldsymbol{E}}{\partial t} \label{maxwell-4}
			\end{align}
		\end{subequations}
		allows us to recognize such an equivalence. But, although this is exactly what we will do throughout this Section, it is convenient to make a small caveat because you, the reader, will see that, instead of replacing the expressions (\ref{relation-rk}) in the equations (\ref{maxwell}), we will replace
		\begin{subequations} \label{campos}
			\begin{align}
				\boldsymbol{E} \left( \boldsymbol{r} , t \right) & = \int \boldsymbol{\mathcal{E}} \left( \boldsymbol{k} , t \right) \hspace*{0.04cm} e^{i \boldsymbol{k} \boldsymbol{\cdot } \boldsymbol{r}} d \boldsymbol{k} \label{campos-1} \quad \textnormal{and} \\
				\boldsymbol{B} \left( \boldsymbol{r} , t \right) & = \int \boldsymbol{\mathcal{B}} \left( \boldsymbol{k} , t \right) \hspace*{0.04cm} e^{i \boldsymbol{k} \boldsymbol{\cdot } \boldsymbol{r}} d \boldsymbol{k} \label{campos-2} \ .
			\end{align}
		\end{subequations}
		And why will we do this? We will do this because, in addition to these expressions (\ref{relation-rk}) and (\ref{campos}) being exactly the same under the identification made in (\ref{eb-identification}), what we want to do here, with the replacement of (\ref{campos}) in Maxwell's equations (\ref{maxwell}), is to rewrite these equations in $ \boldsymbol{k} $-space. And since the electromagnetic fields in the equations (\ref{maxwell}) are expressed as a function of $ \left( \boldsymbol{r} , t \right) $, it is essential that this replacement uses functions of $ \left( \boldsymbol{k} , t \right) $.
		
		Observe that, in order for us to be able to rewrite Maxwell's equations in $ \boldsymbol{k} $-space, it will also be necessary to take
		\begin{subequations} \label{carga-corrente}
			\begin{align}
				\rho \left( \boldsymbol{r} , t \right) & = \int \varrho \left( \boldsymbol{k} , t \right) \hspace*{0.04cm} e^{i \boldsymbol{k} \boldsymbol{\cdot } \boldsymbol{r}} d \boldsymbol{k} \label{campos-3} \quad \textnormal{and} \\
				\boldsymbol{J} \left( \boldsymbol{r} , t \right) & = \int \boldsymbol{\mathcal{J}} \left( \boldsymbol{k} , t \right) \hspace*{0.04cm} e^{i \boldsymbol{k} \boldsymbol{\cdot } \boldsymbol{r}} d \boldsymbol{k} \label{campos-4} \ .
			\end{align}
		\end{subequations}
		As a consequence, one thing that is quite clear here is that, when we substitute, for instance, the expression of the fields (\ref{campos-1}) and (\ref{campos-3}) in the equation (\ref{maxwell-1}), we obtain
		\begin{equation*}
			\boldsymbol{\nabla } _{\boldsymbol{r}} \cdot \int \boldsymbol{\mathcal{E}} \left( \boldsymbol{k} , t \right) e^{i \boldsymbol{k} \cdot \boldsymbol{r}} d \boldsymbol{k} = \int \boldsymbol{\nabla } _{\boldsymbol{r}} \cdot \left[ \boldsymbol{\mathcal{E}} \left( \boldsymbol{k} , t \right) e^{i \boldsymbol{k} \cdot \boldsymbol{r}} \right] d \boldsymbol{k} = \frac{1}{\varepsilon _{0}} \int \varrho \left( \boldsymbol{k} , t \right) e^{i \boldsymbol{k} \cdot \boldsymbol{r}} d \boldsymbol{k} \ .
		\end{equation*}
		Thus, by noting that
		\begin{equation*}
			\boldsymbol{\nabla } _{\boldsymbol{r}} \cdot \left[ \boldsymbol{\mathcal{E}} \left( \boldsymbol{k} , t \right) e^{i \boldsymbol{k} \cdot \boldsymbol{r}} \right] = e^{i \boldsymbol{k} \cdot \boldsymbol{r}} \underbrace{\ \left[ \boldsymbol{\nabla } _{\boldsymbol{r}} \cdot \boldsymbol{\mathcal{E}} \left( \boldsymbol{k} , t \right) \right] \ }_{=0} \ + \ \boldsymbol{\mathcal{E}} \left( \boldsymbol{k} , t \right) \cdot \boldsymbol{\nabla } _{\boldsymbol{r}} \ e^{i \boldsymbol{k} \cdot \boldsymbol{r}} = \left[ i \boldsymbol{k} \cdot \boldsymbol{\mathcal{E}} \left( \boldsymbol{k} , t \right) \right] e^{i \boldsymbol{k} \cdot \boldsymbol{r}} \ ,
		\end{equation*}
		this development ends up showing that
		\begin{equation*}
			i \boldsymbol{k} \cdot \boldsymbol{\mathcal{E}} \left( \boldsymbol{k} , t \right) = \frac{1}{\varepsilon _{0}} \ \varrho \left( \boldsymbol{k} , t \right) \ . 
		\end{equation*}
		Note that, as equation (\ref{maxwell-2}) is analogous to (\ref{maxwell-1}), this last result also allows to conclude that
		\begin{equation*}
			i \boldsymbol{k} \cdot \boldsymbol{\mathcal{B}} \left( \boldsymbol{k} , t \right) = 0 \ .
		\end{equation*}
		
		Now, in relation to what follows, for example, from the replacement of (\ref{campos-1}) and (\ref{campos-2}) in equation (\ref{maxwell-3}), it is immediate to see that
		\begin{equation} 
			\begin{split}
				\boldsymbol{\nabla } _{\boldsymbol{r}} \times \int \boldsymbol{\mathcal{E}} \left( \boldsymbol{k} , t \right) e^{i \boldsymbol{k} \cdot \boldsymbol{r}} d \boldsymbol{k} & = - \frac{\partial }{\partial t} \int \boldsymbol{\mathcal{B}} \left( \boldsymbol{k} , t \right) e^{i \boldsymbol{k} \cdot \boldsymbol{r}} d \boldsymbol{k} \\
				& \Rightarrow \int \boldsymbol{\nabla } _{\boldsymbol{r}} \times \left[  \boldsymbol{\mathcal{E}} \left( \boldsymbol{k} , t \right) e^{i \boldsymbol{k} \cdot \boldsymbol{r}} \right] d \boldsymbol{k} = - \int \frac{\partial }{\partial t} \left[ \boldsymbol{\mathcal{B}} \left( \boldsymbol{k} , t \right) e^{i \boldsymbol{k} \cdot \boldsymbol{r}} \right] d \boldsymbol{k} \ . 
			\end{split} \label{integral-aux-1}
		\end{equation}
		Thus, by noting that
		\begin{equation*}
			\boldsymbol{\nabla } _{\boldsymbol{r}} \times \left[ \boldsymbol{\mathcal{E}} \left( \boldsymbol{k} , t \right) e^{i \boldsymbol{k} \cdot \boldsymbol{r}} \right] = \underbrace{ \ e^{i \boldsymbol{k} \cdot \boldsymbol{r}} \cdot \boldsymbol{\nabla } _{\boldsymbol{r}} \times \boldsymbol{\mathcal{E}} \left( \boldsymbol{k} , t \right) \ } _{=0} + \left( \boldsymbol{\nabla } _{\boldsymbol{r}} \ e^{i \boldsymbol{k} \cdot \boldsymbol{r}} \right) \times \boldsymbol{\mathcal{E}} \left( \boldsymbol{k} , t \right) = i \boldsymbol{k} \ e^{i \boldsymbol{k} \cdot \boldsymbol{r}} \times \boldsymbol{\mathcal{E}} \left( \boldsymbol{k} , t \right) \ ,
		\end{equation*}
		it is not difficult to conclude that (\ref{integral-aux-1}) becomes
		\begin{equation*}
			\int \left[ i \boldsymbol{k} e^{i \boldsymbol{k} \cdot \boldsymbol{r}} \times \boldsymbol{\mathcal{E}} \left( \boldsymbol{k} , t \right) \right] d \boldsymbol{k} = - \int \frac{\partial }{\partial t} \left[ \boldsymbol{\mathcal{B}} \left( \boldsymbol{k} , t \right) e^{i \boldsymbol{k} \cdot \boldsymbol{r}} \right] d \boldsymbol{k} = - \int \left( \frac{\partial \boldsymbol{\mathcal{B}}}{\partial t} \right) e^{i \boldsymbol{k} \cdot \boldsymbol{r}} d\boldsymbol{k} \ ,
		\end{equation*}
		which makes clear that the sufficient condition for this equality to hold is
		\begin{equation*}
			 \boldsymbol{k} \times \boldsymbol{\mathcal{E}} \left( \boldsymbol{k} , t \right) = - i \hspace*{0.04cm} \frac{\partial \boldsymbol{\mathcal{B}}}{\partial t} \ .
		\end{equation*}
		So, since the replacement of (\ref{campos-1}), (\ref{campos-2}) and (\ref{campos-4}) in the equation (\ref{maxwell-4}) also shows us that
		\begin{equation*}
			i \boldsymbol{k} \times \boldsymbol{\mathcal{B}} \left( \boldsymbol{k} , t \right) = \mu _{0} \hspace*{0.04cm} \boldsymbol{\mathcal{J}} \left( \boldsymbol{k} , t \right) + \frac{1}{c^{2}} \frac{\partial \boldsymbol{\mathcal{E}}}{\partial t} \ ,
		\end{equation*}
		it is valid to say that all the equations
		\begin{subequations} \label{remaxwell}
			\begin{align}
				i \boldsymbol{k} \cdot \boldsymbol{\mathcal{E}} \left( \boldsymbol{k} , t \right) & = \frac{1}{\varepsilon _{0}} \hspace*{0.04cm} \varrho \left( \boldsymbol{k} , t \right) \label{remaxwell-1} \ , \\
				i \boldsymbol{k} \cdot \boldsymbol{\mathcal{B}} \left( \boldsymbol{k} , t \right) & = 0 \label{remaxwell-2} \ , \\
				i \boldsymbol{k} \times \boldsymbol{\mathcal{E}} \left( \boldsymbol{k} , t \right) & = - \frac{\partial \boldsymbol{\mathcal{B}}}{\partial t} \label{remaxwell-3} \ \ \textnormal{and} \\
				i \boldsymbol{k} \times \boldsymbol{\mathcal{B}} \left( \boldsymbol{k} , t \right) & = \mu _{0} \hspace*{0.04cm} \boldsymbol{\mathcal{J}} \left( \boldsymbol{k} , t \right) + \frac{1}{c^{2}} \frac{\partial \boldsymbol{\mathcal{E}}}{\partial t} \label{remaxwell-4}
			\end{align}
		\end{subequations}
		are equivalent to the same Maxwell's equations (\ref{maxwell}). Observe that, by replacing (\ref{campos-3}) and (\ref{campos-4}) into the continuity equation \cite{jackson}
		\begin{equation*}
			\boldsymbol{\nabla } _{\boldsymbol{r}} \cdot \boldsymbol{J} + \frac{\partial \rho }{\partial t} = 0 \ ,
		\end{equation*}
		manipulations analogous to the ones we did above lead us to
		\begin{equation}
			i \boldsymbol{k} \cdot \boldsymbol{\mathcal{J}} \left( \boldsymbol{k} , t \right) + \frac{\partial \varrho }{\partial t} = 0 \ . \label{continuidade}
		\end{equation} 
	
		\subsection{Electromagnetic wave equations written in $ \boldsymbol{k} $-space}
		
			From the results we have just derived, it is not difficult to see that, by manipulating these equations (\ref{remaxwell}), we obtain two differential equations that, for example, show us that the electromagnetic field behaves like a harmonic oscillator in $ \boldsymbol{k} $-space. And, in order to obtain the first of them, it is enough to notice that equation (\ref{remaxwell-3}) shows us that
			\begin{eqnarray}
				i \boldsymbol{k} \times \left[ \boldsymbol{k} \times \boldsymbol{\mathcal{E}} \left( \boldsymbol{k} , t \right) \right] & = &  \left[ i \boldsymbol{k} \cdot \boldsymbol{\mathcal{E}} \left( \boldsymbol{k} , t \right) \right] \boldsymbol{k} - i \left( \boldsymbol{k} \cdot \boldsymbol{k} \right) \boldsymbol{\mathcal{E}} \left( \boldsymbol{k} , t \right) \notag \\
				& = & \left[ \frac{1}{\varepsilon _{0}} \hspace*{0.04cm} \varrho \left( \boldsymbol{k} , t \right) \right] \boldsymbol{k} - i k^{2} \hspace*{0.04cm} \boldsymbol{\mathcal{E}} \left( \boldsymbol{k} , t \right) = - \boldsymbol{k} \times \frac{\partial \boldsymbol{\mathcal{B}}}{\partial t} \ , \label{manipulation-1}
			\end{eqnarray}
			where $ k = \left\vert \boldsymbol{k} \right\vert $. After all, since equation (\ref{remaxwell-4}) also shows that			
			\begin{equation*}
				\boldsymbol{k} \times \frac{\partial \boldsymbol{\mathcal{B}}}{\partial t} =  \frac{\partial }{\partial t} \left[ \boldsymbol{k} \times \boldsymbol{\mathcal{B}} \left( \boldsymbol{k} , t \right) \right] = - i \mu _{0} \frac{\partial \boldsymbol{\mathcal{J}}}{\partial t} - \frac{i}{c^{2}} \frac{\partial ^{2} \boldsymbol{\mathcal{E}}}{\partial t^{2}} \ ,
			\end{equation*}
			the substitution of this last result into (\ref{manipulation-1}) makes it quite clear that
			\begin{equation}
				\left[ \frac{\partial ^{2}}{\partial t^{2}} + \left( ck \right) ^{2} \right] \boldsymbol{\mathcal{E}} \left( \boldsymbol{k} , t \right) = - \frac{1}{\varepsilon _{0}} \left\{ \left[ i c^{2} \varrho \left( \boldsymbol{k} , t \right) \right] \boldsymbol{k} + \frac{\partial \boldsymbol{\mathcal{J}}}{\partial t} \right\} \ . \label{pre-electric-equation}
			\end{equation}
			Now, in order to obtain the second differential equation, it is enough to observe that the manipulation of the equation (\ref{remaxwell-4}) shows us that	
			\begin{eqnarray}
				i \boldsymbol{k} \times \left[ \boldsymbol{k} \times \boldsymbol{\mathcal{B}} \left( \boldsymbol{k} , t \right) \right] & = & \underbrace{ \ \left[ \boldsymbol{k} \cdot \boldsymbol{\mathcal{B}} \left( \boldsymbol{k} , t \right) \right] \ } _{=0} \boldsymbol{k} - \left( \boldsymbol{k} \cdot \boldsymbol{k} \right) \boldsymbol{\mathcal{B}} \left( \boldsymbol{k} , t \right) \notag \\
				& = & - ik^{2} \boldsymbol{\mathcal{B}} \left( \boldsymbol{k} , t \right) = \mu _{0} \left[ \boldsymbol{k} \times \boldsymbol{\mathcal{J}} \left( \boldsymbol{k} , t \right) \right] + \frac{1}{c^{2}} \left( \boldsymbol{k} \times \frac{\partial \boldsymbol{\mathcal{E}}}{\partial t} \right) \ . \label{manipulation-2}
			\end{eqnarray}
			In this way, since the equation (\ref{remaxwell-3}) also shows us that	
			\begin{equation*}
				\boldsymbol{k} \times \frac{\partial \boldsymbol{\mathcal{E}}}{\partial t} =  \frac{\partial }{\partial t} \left[ \boldsymbol{k} \times \boldsymbol{\mathcal{E}} \left( \boldsymbol{k} , t \right) \right] = i \frac{\partial ^{2} \boldsymbol{\mathcal{B}}}{\partial t^{2}} \ ,
			\end{equation*}
			it is precisely the substitution of this last result into (\ref{manipulation-2}) that leads us to
			\begin{equation}
				\left[ \frac{\partial ^{2}}{\partial t^{2}} + \left( ck \right) ^{2} \right] \boldsymbol{\mathcal{B}} \left( \boldsymbol{k} , t \right) = - i \mu _{0} \left[ \boldsymbol{k} \times \boldsymbol{\mathcal{J}} \left( \boldsymbol{k} , t \right) \right] \ . \label{pre-magnetic-equation}
			\end{equation}
			
			Clearly, these differential equations (\ref{pre-electric-equation}) and (\ref{pre-magnetic-equation}) are telling us that the electromagnetics fields $ \left( \boldsymbol{\mathcal{E}} \left( \boldsymbol{k} , t \right) , \boldsymbol{\mathcal{B}} \left( \boldsymbol{k} , t \right) \right) $, which satisfy Maxwell's equations (\ref{remaxwell}), can also be recognized as electromagnetic waves whose natural frequency of oscillation is given by $ \omega \left( \boldsymbol{k} \right) = c \left\vert \boldsymbol{k} \right\vert = ck $. But, as much as the right-hand sides of (\ref{pre-electric-equation}) and (\ref{pre-magnetic-equation}) characterize these electromagnetic fields as a kind of forced harmonic oscillators, it is interesting to highlight that there is a special case, which is of extreme interest in Physics, where the relationship between $ \boldsymbol{\mathcal{J}} \left( \boldsymbol{k} , t \right) $ and $ \boldsymbol{\mathcal{E}} \left( \boldsymbol{k} , t \right) $ is ohmic \cite{drude}: i.e., where this relationship is such that
			\begin{equation}
				\boldsymbol{\mathcal{J}} \left( \boldsymbol{k} , t \right) = \sigma \hspace*{0.04cm} \boldsymbol{\mathcal{E}} \left( \boldsymbol{k} , t \right) \ , \label{ohmic}
			\end{equation}
			where the electrical conductivity $ \sigma $ can be considered constant. And since this ohmic relationship shows that
			\begin{equation*}
				\frac{\partial \boldsymbol{\mathcal{J}}}{\partial t} = \sigma \hspace*{0.04cm} \frac{\partial \boldsymbol{\mathcal{E}}}{\partial t} \quad \textnormal{and} \quad \boldsymbol{k} \times \boldsymbol{\mathcal{J}} \left( \boldsymbol{k} , t \right) = \sigma \hspace*{0.04cm} \left[ \boldsymbol{k} \times \boldsymbol{\mathcal{E}} \left( \boldsymbol{k} , t \right) \right] = i \sigma \hspace*{0.04cm} \frac{\partial \boldsymbol{\mathcal{B}}}{\partial t} \ ,
			\end{equation*}
			the substitution of these last results into (\ref{pre-electric-equation}) and (\ref{pre-magnetic-equation}) respectively  leads us to
			\begin{subequations} \label{dif-eq}
				\begin{align}
					\left[ \frac{\partial ^{2}}{\partial t^{2}} + \frac{\sigma }{\varepsilon _{0}} \frac{\partial }{\partial t} + \left( ck \right) ^{2} \right] \boldsymbol{\mathcal{E}} \left( \boldsymbol{k} , t \right) & = - \left[ \frac{i c^{2}}{\varepsilon _{0}} \varrho \left( \boldsymbol{k} , t \right) \right] \boldsymbol{k} \quad \textnormal{and} \label{dif-eq-e}\\
					\left[ \frac{\partial ^{2}}{\partial t^{2}} + \frac{\sigma }{\varepsilon _{0}} \frac{\partial }{\partial t} + \left( ck \right) ^{2} \right] \boldsymbol{\mathcal{B}} \left( \boldsymbol{k} , t \right) & = \boldsymbol{0} \label{dif-eq-b} \ ,
				\end{align}
			\end{subequations}
			which makes it quite clear that, in this ohmic special case, the corresponding electromagnetic fields describe damped harmonic oscillations, whose damping factor is given by $ \gamma = \sigma / \varepsilon _{0} $.
					
		\subsection{\label{subsec02}A remark on the components of the electric and magnetic fields}
	
			In accordance with these equations (\ref{dif-eq}), one of the cases where this damping factor disappears is in free space: i.e., a case where, because there are also no electric charges and currents in the medium where the electromagnetic fields are defined, these equations (\ref{dif-eq-b}) reduce to
			\begin{subequations} \label{dif-eq-vacuo}
				\begin{align}
					\left[ \frac{\partial ^{2}}{\partial t^{2}} + \left( ck \right) ^{2} \right] \boldsymbol{\mathcal{E}} \left( \boldsymbol{k} , t \right) & = \boldsymbol{0} \quad \textnormal{and} \label{dif-eq-vacuo-e} \\
					\left[ \frac{\partial ^{2}}{\partial t^{2}} + \left( ck \right) ^{2} \right] \boldsymbol{\mathcal{B}} \left( \boldsymbol{k} , t \right) & = \boldsymbol{0} \label{dif-eq-vacuo-b} \ 
				\end{align}
			\end{subequations}
			But, before we pay attention to this free space case, whose analysis is crucial in view of the proposal of this paper, it is important to remember that, as $ \boldsymbol{E} \left( \boldsymbol{r} , t \right) $ and $ \boldsymbol{B} \left( \boldsymbol{r} , t \right) $ describe an electromagnetic field, these two functions must be real. And remembering this is very significant because this reality condition, being such that
			\begin{equation}
				\boldsymbol{E} \left( \boldsymbol{r} , t \right) = \boldsymbol{E} ^{\ast } \left( \boldsymbol{r} , t \right) \quad \textnormal{and} \quad \boldsymbol{B} \left( \boldsymbol{r} , t \right) = \boldsymbol{B} ^{\ast } \left( \boldsymbol{r} , t \right) \ , \label{realidade-campos}
			\end{equation}
			makes it clear that $ \boldsymbol{\mathcal{E}} \left( \boldsymbol{k} , t \right) $ and $ \boldsymbol{\mathcal{B}} \left( \boldsymbol{k} , t \right) $ need to satisfy some constraints. Note that such constraints can be obtained as long as we observe not only that 
			\begin{equation}
				\boldsymbol{E} ^{\ast } \left( \boldsymbol{r} , t \right) = \int \boldsymbol{\mathcal{E}} ^{\ast } \left( \boldsymbol{k} , t \right) e^{-i \boldsymbol{k} \cdot \boldsymbol{r}} d \boldsymbol{k} \quad \textnormal{and} \quad \boldsymbol{B} ^{\ast } \left( \boldsymbol{r} , t \right) = \int \boldsymbol{\mathcal{B}} ^{\ast } \left( \boldsymbol{k} , t \right) e^{-i \boldsymbol{k} \cdot \boldsymbol{r}} d \boldsymbol{k} \ , \label{conjugated-campos}
			\end{equation}
			but also that a change of variable $ \boldsymbol{k} \rightarrow - \boldsymbol{k} $ leads us to
			\begin{eqnarray}
				\boldsymbol{E} \left( \boldsymbol{r} , t \right) \negthickspace & = & \negthickspace - \int \boldsymbol{\mathcal{E}} \left( - \boldsymbol{k} , t \right) e^{-i \boldsymbol{k} \cdot \boldsymbol{r}} d \boldsymbol{k} = \int \boldsymbol{\mathcal{E}} \left( - \boldsymbol{k} , t \right) e^{-i \boldsymbol{k} \cdot \boldsymbol{r}} d \boldsymbol{k} \ \ \textnormal{and} \notag \\ 
				\boldsymbol{B} \left( \boldsymbol{r} , t \right) \negthickspace & = & \negthickspace - \int \boldsymbol{\mathcal{B}} \left( - \boldsymbol{k} , t \right) e^{-i \boldsymbol{k} \cdot \boldsymbol{r}} d \boldsymbol{k} = \int \boldsymbol{\mathcal{B}} \left( - \boldsymbol{k} , t \right) e^{-i \boldsymbol{k} \cdot \boldsymbol{r}} d \boldsymbol{k} \ . \label{alternative-campos}
			\end{eqnarray}
			After all, by replacing (\ref{conjugated-campos}) and (\ref{alternative-campos}) in the reality conditions (\ref{realidade-campos}), it is not difficult to conclude that these constraints are
			\begin{equation}
				\boldsymbol{\mathcal{E}} ^{\ast } \left( \boldsymbol{k} , t \right) = \boldsymbol{\mathcal{E}} \left( - \boldsymbol{k} , t \right) \quad \textnormal{and} \quad \boldsymbol{\mathcal{B}} ^{\ast } \left( \boldsymbol{k} , t \right) = \boldsymbol{\mathcal{B}} \left( - \boldsymbol{k} , t \right) \ . \label{realidade-coordenadas}
			\end{equation}
	
	\section{\label{sec03}Getting a Schr\"{o}dinger equation in free space}
		
		Given this result (\ref{realidade-coordenadas}), it is not difficult to see, for instance, that, with the help of another function $ \boldsymbol{\varphi } \left( \boldsymbol{k} , t \right) $, the function $ \boldsymbol{\mathcal{E}} \left( \boldsymbol{k} , t \right) $ can be expressed as
		\begin{equation}
			\boldsymbol{\mathcal{E}} \left( \boldsymbol{k} , t \right) = n \left( k \right) \bigl[ \boldsymbol{\varphi } \left( \boldsymbol{k} , t \right) + \boldsymbol{\varphi } ^{\ast } \left( - \boldsymbol{k} , t \right) \bigr] \label{fabulous-electric} \ ,
		\end{equation}
		where $ n \left( k \right) $ is a function that only depends on the modulus $ k = \left\vert \boldsymbol{k} \right\vert $. And in order to get a similar expression for $ \boldsymbol{\mathcal{B}} \left( \boldsymbol{k} , t \right) $, the first thing we can do is notice, for instance, that the equation (\ref{manipulation-2}) leads us to
		\begin{equation*}
			\boldsymbol{\mathcal{B}} \left( \boldsymbol{k} , t \right) = i \boldsymbol{k} \times \left[ \frac{\sigma \mu _{0}}{k^{2}} + \frac{1}{\left( ck \right) ^{2}} \frac{\partial }{\partial t} \right] \boldsymbol{\mathcal{E}} \left( \boldsymbol{k} , t \right) \ .
		\end{equation*}	
		After all, by noticing that, when $ \sigma = 0 $, this last result reduces to
		\begin{equation}
			\boldsymbol{\mathcal{B}} \left( \boldsymbol{k} , t \right) = \frac{i}{\left( ck \right) ^{2}} \ \boldsymbol{k} \times \boldsymbol{\dot{\mathcal{E}}} \left( \boldsymbol{k} , t \right) \label{magnetic-derivative-relation} \ ,
		\end{equation}		
		this is exactly what allows us to conclude that, in free space, $ \boldsymbol{\mathcal{B}} \left( \boldsymbol{k} , t \right) $ can be expressed, at least, as a function of the (first order) time derivative of $ \boldsymbol{\varphi } \left( \boldsymbol{k} , t \right) $ because (\ref{fabulous-electric}) shows us that
		\begin{equation}
			\boldsymbol{\dot{\mathcal{E}}} \left( \boldsymbol{k} , t \right) = n \left( k \right) \bigl[ \boldsymbol{\dot{\varphi }} \left( \boldsymbol{k} , t \right) + \boldsymbol{\dot{\varphi }} ^{\ast } \left( - \boldsymbol{k} , t \right) \bigr] \ . \label{e-derivadas-f}
		\end{equation}
		Therefore, as the development of (\ref{dif-eq-vacuo-e}) shows us that
		\begin{equation*}
			\left( \frac{\partial }{\partial t} + i ck \right) \boldsymbol{\mathcal{E}} \left( \boldsymbol{k} , t \right) = \boldsymbol{0} \quad \textnormal{or} \quad \left( \frac{\partial }{\partial t} - i ck \right) \boldsymbol{\mathcal{E}} \left( \boldsymbol{k} , t \right) = \boldsymbol{0} \ ,
		\end{equation*}			
		the expressions (\ref{fabulous-electric}) and (\ref{e-derivadas-f}) make it clear that solving this equation (\ref{dif-eq-vacuo-e}) means making explicit the relationship that exists between $ \boldsymbol{\varphi } \left( \boldsymbol{k} , t \right) $ and $ \boldsymbol{\dot{\varphi }} \left( \boldsymbol{k} , t \right) $ in free space: more specifically, a relationship that, in free space, can be chosen, for example, as
		\begin{equation}
			ck \hspace*{0.04cm} \boldsymbol{\varphi } \left( \boldsymbol{k} , t \right) = i \frac{\partial \boldsymbol{\varphi }}{\partial t} \label{f-first-order}
		\end{equation}
		which, when substituted in (\ref{magnetic-derivative-relation}), allows us to conclude that
		\begin{equation}
			\boldsymbol{\mathcal{B}} \left( \boldsymbol{k} , t \right) =  \frac{n \left( k \right) }{ck} \ \boldsymbol{k} \times \bigl[ \boldsymbol{\varphi } \left( \boldsymbol{k} , t \right) - \boldsymbol{\varphi } ^{\ast } \left( - \boldsymbol{k} , t \right) \bigr] \ . \label{fabulous-electric-derivative}
		\end{equation}
	
		In view of what we have done to obtain Maxwell's equations in $ \boldsymbol{k} $-space
		\begin{subequations} \label{remaxwell-free}
			\begin{align}
				i \boldsymbol{k} \cdot \boldsymbol{\mathcal{E}} \left( \boldsymbol{k} , t \right) & = 0 \label{remaxwell-free-1} \ , \\
				i \boldsymbol{k} \cdot \boldsymbol{\mathcal{B}} \left( \boldsymbol{k} , t \right) & = 0 \label{remaxwell-free-2} \ , \\
				i \boldsymbol{k} \times \boldsymbol{\mathcal{E}} \left( \boldsymbol{k} , t \right) & = - \frac{\partial \boldsymbol{\mathcal{B}}}{\partial t} \label{remaxwell-free-3} \ \ \textnormal{and} \\
				i \boldsymbol{k} \times \boldsymbol{\mathcal{B}} \left( \boldsymbol{k} , t \right) & =  \frac{1}{c^{2}} \frac{\partial \boldsymbol{\mathcal{E}}}{\partial t} \label{remaxwell-free-4}
			\end{align}
		\end{subequations}
		in free space, it is not wrong to acknowledge that (\ref{f-first-order}) was obtained by manipulating only the (\ref{remaxwell-free-3}) and (\ref{remaxwell-free-4}), and not the (\ref{remaxwell-free-1}) and (\ref{remaxwell-free-2}). However, by substituting either of these expressions (\ref{fabulous-electric}) or (\ref{fabulous-electric-derivative}) into Maxwell's equations (\ref{remaxwell-free-1}) and (\ref{remaxwell-free-2}), what we get is
		\begin{equation}
			\boldsymbol{k} \boldsymbol{\cdot } \boldsymbol{\varphi } \left( \boldsymbol{k} , t \right) = \sum ^{3} _{\beta = 1} k_{\beta } \cdot \varphi _{\beta } \left( \boldsymbol{k} , t \right) = 0 \ , \label{f-perpendicular}
		\end{equation}
		which makes clear that another way to rewrite the Maxwell's equations (\ref{maxwell}) in free space is through
		\begin{equation}
			ck \hspace*{0.04cm} \boldsymbol{\varphi } \left( \boldsymbol{k} , t \right) = i \hspace*{0.04cm} \frac{\partial \boldsymbol{\varphi }}{\partial t} \quad \textnormal{and} \quad \boldsymbol{k} \boldsymbol{\cdot } \boldsymbol{\varphi } \left( \boldsymbol{k} , t \right) = 0 \ . \label{sistema-fk}
		\end{equation}		
		After all, note that solving this new system (\ref{sistema-fk}) means determining the function $ \boldsymbol{\varphi } \left( \boldsymbol{k} , t \right) $ that defines the functions (\ref{fabulous-electric}) and (\ref{fabulous-electric-derivative}), and consequently determining the electric $ \boldsymbol{E} \left( \boldsymbol{r} , t \right) $ and magnetic $ \boldsymbol{B} \left( \boldsymbol{r} , t \right) $ fields that satisfy Maxwell's equations (\ref{maxwell}) in free space.
		
		Yet, another thing we need to note here is that, due to the linear independence of the equations in the system (\ref{sistema-fk}), this is not the only way to express the Maxwell's equations in free space in terms of $ \boldsymbol{\varphi } \left( \boldsymbol{k} , t \right) $: another way is by \cite{elon}
		\begin{itemize}
			\item[\textbf{(i)}] multiplying the equation (\ref{f-first-order}) by a real constant $ a $,
			\begin{equation*}
				ack \hspace*{0.04cm} \boldsymbol{\varphi } \left( \boldsymbol{k} , t \right) = ack \hspace*{0.04cm} \left[ \sum ^{3} _{\beta = 1} \varphi _{\beta } \left( \boldsymbol{k} , t \right) \right] = ack \hspace*{0.04cm} \left[ \sum ^{3} _{\alpha = 1} \sum ^{3} _{\beta = 1} \delta _{\alpha \beta } \varphi _{\beta } \left( \boldsymbol{k} , t \right) \right] = i a \frac{\partial \boldsymbol{\varphi }}{\partial t} \ ,
			\end{equation*}
			\item[\textbf{(ii)}] multiplying the equation (\ref{f-perpendicular}) by the vector $ c \boldsymbol{k} / k $ previously multiplied by the same constant $ a $ mentioned in \textbf{(i)},
			\begin{equation*}
				\frac{ac \hspace*{0.04cm} \boldsymbol{k}}{k} \left[ \boldsymbol{k} \boldsymbol{\cdot } \boldsymbol{\varphi } \left( \boldsymbol{k} , t \right) \right] = \frac{ac}{k} \left( \sum ^{3} _{\alpha = 1} k_{\alpha } \right) \left[ \sum ^{3} _{\beta = 1} k_{\beta } \cdot \varphi _{\beta } \left( \boldsymbol{k} , t \right) \right] = 0 \ , \ \ \textnormal{and}
			\end{equation*} 
			\item[\textbf{(iii)}] subtracting the result \textbf{(ii)} from the \textbf{(i)},
			\begin{equation}
				\sum ^{3} _{\alpha = 1} \sum ^{3} _{\beta = 1} ack \left( \delta _{\alpha \beta } - \frac{k_{\alpha } k_{\beta }}{k^{2}} \right) \varphi _{\beta } \left( \boldsymbol{k} , t \right) = i a \frac{\partial \boldsymbol{\varphi }}{\partial t} \ , \label{pre-schrodinger}
			\end{equation}
		\end{itemize}
		And, by taking $ a $ as $ \hbar $ in the last result, this brings us to a new equation
		\begin{equation}
			\mathbb{H} \hspace*{0.04cm} \boldsymbol{\varphi } \left( \boldsymbol{k} , t \right) = i \hbar \frac{\partial \boldsymbol{\varphi }}{\partial t} \label{maxwell-schrodinger}
		\end{equation}
		that is very peculiar, since $ \mathbb{H} \hspace*{0.04cm} \boldsymbol{\varphi } \left( \boldsymbol{k} , t \right) $ is a vector whose components are
		\begin{equation}
			\mathbb{H} \hspace*{0.04cm} \varphi _{\alpha } \left( \boldsymbol{k} , t \right) = \sum ^{3} _{\beta = 1} \mathbb{H} _{\alpha \beta } \cdot \varphi _{\beta } \left( \boldsymbol{k} , t \right) \ , \quad \textnormal{where} \quad \mathbb{H} _{\alpha \beta } = \hbar ck \left( \delta _{\alpha \beta } - \frac{k_{\alpha } k_{\beta }}{k^{2}} \right) \ . \label{maxwell-schrodinger-componentes}
		\end{equation}
		In other words, this new development, which has just been described by items \textbf{(i)}, \textbf{(ii)} and \textbf{(iii)}, is leading us to a new system
		\begin{equation}
			\mathbb{H} \hspace*{0.04cm} \boldsymbol{\varphi } \left( \boldsymbol{k} , t \right) = i \hbar \frac{\partial \boldsymbol{\varphi }}{\partial t} \quad \textnormal{and} \quad \boldsymbol{k} \boldsymbol{\cdot } \boldsymbol{\varphi } \left( \boldsymbol{k} , t \right) = 0 \label{sistema-maxwell-schrodinger}
		\end{equation}
		that covers the same content as Maxwell's equations in free space, but with one of its equations taking the same form as the Schr\"{o}dinger equation. 
		
		\subsection{\label{subsec031}Some considerations on the energy}
		
			Although it is already quite tempting to claim that (\ref{maxwell-schrodinger}) is a Schr\"{o}dinger equation, we still cannot do this without first checking whether, for example, $ \mathbb{H} $ can be interpreted as a Hamiltonian operator. And since every Hamiltonian operator measures the energy of the system associated with it, a good way to assess whether $ \mathbb{H} $ really can be interpreted as such is by evaluating not only whether there is any relationship between it and the energy
			\begin{equation}
				\mathbb{E} = \frac{\varepsilon _{0}}{2} \int \left[ \boldsymbol{E} ^{2} \left( \boldsymbol{r} , t \right) + c^{2} \boldsymbol{B} ^{2} \left( \boldsymbol{r} , t \right) \right] d \boldsymbol{r} \label{foton-energy}
			\end{equation}
			of electromagnetic waves that propagate through free space, but also whether this energy (\ref{foton-energy}) leads to a known result, which is the energy of a photon.
			
			One of the things we can do to make this assessment is, for example, to deal with the result
			\begin{equation}
				\int e^{i \left( \boldsymbol{k + k^{\prime }} \right) \boldsymbol{r}} \ d \boldsymbol{r} = \left( 2 \pi \right) ^{3} \delta \left( \boldsymbol{k + k^{\prime }} \right) \ . \label{delta-function}
			\end{equation}
			After all, when we substitute (\ref{campos-1}) and (\ref{campos-2}) into equation (\ref{foton-energy}), it allows us to observe that
			\begin{eqnarray}
				\mathbb{E} \negthickspace & = & \negthickspace \frac{\varepsilon _{0}}{2} \int \int \left[ \boldsymbol{\mathcal{E}} \left( \boldsymbol{k} , t \right) \cdot \boldsymbol{\mathcal{E}} \left( \boldsymbol{k^{\prime }} , t \right) + c^{2} \boldsymbol{\mathcal{B}} \left( \boldsymbol{k} , t \right) \cdot \boldsymbol{\mathcal{B}} \left( \boldsymbol{k^{\prime }} , t \right) \right] \int e^{i \left( \boldsymbol{k + k^{\prime }} \right) \boldsymbol{r}} \ d \boldsymbol{r} \hspace*{0.04cm} d \boldsymbol{k} \hspace*{0.04cm} d \boldsymbol{k^{\prime }} \notag \\
				& = & \negthickspace \frac{\left( 2 \pi \right) ^{3} \varepsilon _{0}}{2} \int \left[ \boldsymbol{\mathcal{E}} \left( \boldsymbol{k} , t \right) \cdot \boldsymbol{\mathcal{E}} \left( - \boldsymbol{k} , t \right) + c^{2} \boldsymbol{\mathcal{B}} \left( \boldsymbol{k} , t \right) \cdot \boldsymbol{\mathcal{B}} \left( - \boldsymbol{k} , t \right) \right] d \boldsymbol{k} \ . \label{preliminar-energy}
			\end{eqnarray}
			Observe that, with the help of (\ref{fabulous-electric}), the first term of this integral (\ref{preliminar-energy}) can be developed as
			\begin{eqnarray}
				\boldsymbol{\mathcal{E}} \left( \boldsymbol{k} , t \right) \cdot \boldsymbol{\mathcal{E}} \left( - \boldsymbol{k} , t \right) \negthickspace & = & \negthickspace \left[ n \left( k \right) \right] ^{2} \bigl[ \boldsymbol{\varphi } \left( \boldsymbol{k} , t \right) \cdot \boldsymbol{\varphi } \left( - \boldsymbol{k} , t \right) + \boldsymbol{\varphi } \left( \boldsymbol{k} , t \right) \cdot \boldsymbol{\varphi } ^{\ast } \left( \boldsymbol{k} , t \right) \bigr] \notag \\
				& & \negthickspace + \left[ n \left( k \right) \right] ^{2} \bigl[ \boldsymbol{\varphi } ^{\ast } \left( - \boldsymbol{k} , t \right) \cdot \boldsymbol{\varphi } \left( - \boldsymbol{k} , t \right) + \boldsymbol{\varphi } ^{\ast } \left( - \boldsymbol{k} , t \right) \cdot \boldsymbol{\varphi } ^{\ast } \left( \boldsymbol{k} , t \right) \bigr] \label{e-prod}
			\end{eqnarray}
			and that, with the help of (\ref{fabulous-electric-derivative}), it is not difficult to see that
			\begin{equation*}
				\boldsymbol{\mathcal{B}} \left( \boldsymbol{k} , t \right) \cdot \boldsymbol{\mathcal{B}} \left( - \boldsymbol{k} , t \right) = \frac{1}{\left( ck \right) ^{4}} \left( \boldsymbol{k} \cdot \boldsymbol{k} \right) \left[ \boldsymbol{\dot{\mathcal{E}}} \left( \boldsymbol{k} , t \right) \cdot \boldsymbol{\dot{\mathcal{E}}} \left( - \boldsymbol{k} , t \right) \right] + \frac{1}{\left( ck \right) ^{4}} \left[ \boldsymbol{\dot{\mathcal{E}}} \left( \boldsymbol{k} , t \right) \cdot \boldsymbol{k} \right] \left[ \boldsymbol{k} \cdot \boldsymbol{\dot{\mathcal{E}}} \left( - \boldsymbol{k} , t \right) \right] \ .
			\end{equation*}
			Thus, given that the equation (\ref{remaxwell-free-1}) is such that
			\begin{equation*}
				\boldsymbol{k} \cdot \boldsymbol{\mathcal{E}} \left( \boldsymbol{k} , t \right) = 0 \ \Rightarrow \ \frac{\partial }{\partial t} \left[ \boldsymbol{k} \cdot \boldsymbol{\mathcal{E}} \left( \boldsymbol{k} , t \right) \right] = \boldsymbol{k} \cdot \frac{\partial \boldsymbol{\mathcal{E}}}{\partial t} = 0
			\end{equation*}
			and, therefore, leads to
			\begin{eqnarray}
				\boldsymbol{\mathcal{B}} \left( \boldsymbol{k} , t \right) \cdot \boldsymbol{\mathcal{B}} \left( - \boldsymbol{k} , t \right) \negthickspace & = & \negthickspace \frac{1}{\left( ck \right) ^{2}} \left[ \boldsymbol{\dot{\mathcal{E}}} \left( \boldsymbol{k} , t \right) \cdot \boldsymbol{\dot{\mathcal{E}}} \left( - \boldsymbol{k} , t \right) \right] \notag \\
				\negthickspace & = & \negthickspace - \frac{\left[ n \left( k \right) \right] ^{2}}{c^{2}} \ \bigl[ \boldsymbol{\varphi } \left( \boldsymbol{k} , t \right) \cdot \boldsymbol{\varphi } \left( - \boldsymbol{k} , t \right) - \boldsymbol{\varphi } \left( \boldsymbol{k} , t \right) \cdot \boldsymbol{\varphi } ^{\ast } \left( \boldsymbol{k} , t \right) \bigr] \notag \\
				\negthickspace & + & \negthickspace \frac{\left[ n \left( k \right) \right] ^{2}}{c^{2}} \ \bigl[ \boldsymbol{\varphi } ^{\ast } \left( - \boldsymbol{k} , t \right) \cdot \boldsymbol{\varphi } \left( - \boldsymbol{k} , t \right) - \boldsymbol{\varphi } ^{\ast } \left( - \boldsymbol{k} , t \right) \cdot \boldsymbol{\varphi } ^{\ast } \left( \boldsymbol{k} , t \right) \bigr] \ , \label{b-prod}
			\end{eqnarray}
			it is exactly this last result that, together with (\ref{e-prod}), shows us that this energy (\ref{preliminar-energy}) can be rewritten as
			\begin{equation*}
				\mathbb{E} = \left( 2 \pi \right) ^{3} \varepsilon _{0} \left[ \ \int \left[ n \left( k \right) \right] ^{2} \boldsymbol{\varphi } \left( \boldsymbol{k} , t \right) \cdot \boldsymbol{\varphi } ^{\ast } \left( \boldsymbol{k} , t \right) \ d \boldsymbol{k} \ + \int \left[ n \left( k \right) \right] ^{2} \boldsymbol{\varphi } ^{\ast } \left( - \boldsymbol{k} , t \right) \cdot \boldsymbol{\varphi } \left( - \boldsymbol{k} , t \right) \ d \boldsymbol{k} \ \right] \ .
			\end{equation*}
			And why is interesting to obtain such a result? Because, when we perform a change $ \boldsymbol{k} \rightarrow - \boldsymbol{k} $ in the last integral, this expression, which we got for $ \mathbb{E} $, can be further reduced to
			\begin{equation}
				\mathbb{E} = 2 \left( 2 \pi \right) ^{3} \varepsilon _{0} \int \left[ n \left( k \right) \right] ^{2} \boldsymbol{\varphi } ^{\ast } \left( \boldsymbol{k} , t \right) \cdot \boldsymbol{\varphi } \left( \boldsymbol{k} , t \right) \ d \boldsymbol{k} \ , \label{foton-energy-nd}
			\end{equation}
			which makes this task of evaluating whether (\ref{maxwell-schrodinger}) actually corresponds to a Schr\"{o}dinger equation become easier. And as the equation (\ref{f-first-order}) also helps us to recognize, for instance, that 
			\begin{equation}
				\frac{\partial ^{2} \boldsymbol{\varphi }}{\partial t^{2}} = - ck \left( i \frac{\partial \boldsymbol{\varphi }}{\partial t} \right) = - \left( ck \right) ^{2} \boldsymbol{\varphi } \left( \boldsymbol{k} , t \right) \ \Rightarrow \ \frac{\partial ^{2} \boldsymbol{\varphi }}{\partial t^{2}} + \left( ck \right) ^{2} \boldsymbol{\varphi } \left( \boldsymbol{k} , t \right) = 0 \ , \label{wave-equation-h-aux}
			\end{equation}
			this makes it very clear that, in this region, where there are no electric charges or currents, such an evaluation can be performed by considering
			\begin{equation}
				\boldsymbol{\varphi } \left( \boldsymbol{k} , t \right) = \boldsymbol{\varphi } _{0} \hspace*{-0.04cm} \left( \boldsymbol{k} \right) e^{-i \omega t} \label{wave-function}
			\end{equation}
			as solution of the equation (\ref{maxwell-schrodinger}), where $ \boldsymbol{\varphi } _{0} \hspace*{-0.04cm} \left( \boldsymbol{k} \right) = \boldsymbol{\varphi } \left( \boldsymbol{k} , 0 \right)$ and $ \omega = ck $ \footnote{Note that dealing with this solution (\ref{wave-function}) makes perfect sense since, for example, it is fully consistent with the fact that, by taking $ \omega \left( \boldsymbol{k} \right) = \omega $, the expressions (\ref{relation-rk}) and (\ref{campos}) are exactly the same under the identification made in (\ref{eb-identification}).}. After all, since the substitution of this monochromatic solution (\ref{wave-function}) in the equation (\ref{maxwell-schrodinger}) shows us that
			\begin{equation}
				\mathbb{H} \hspace*{0.04cm} \boldsymbol{\varphi } \left( \boldsymbol{k} , t \right) = \hbar \omega \ \boldsymbol{\varphi } \left( \boldsymbol{k} , t \right) \ , \label{eigen-equation}
			\end{equation}
			it is precisely this that makes it quite clear that
			\begin{equation}
				\int \boldsymbol{\varphi } ^{\ast } \left( \boldsymbol{k} , t \right) \cdot \left[ \mathbb{H} \hspace*{0.04cm} \boldsymbol{\varphi } \left( \boldsymbol{k} , t \right) \right] \ d \boldsymbol{k} = \hbar \omega \int \boldsymbol{\varphi } ^{\ast } \left( \boldsymbol{k} , t \right) \cdot \boldsymbol{\varphi } \left( \boldsymbol{k} , t \right) \ d \boldsymbol{k} \ . \label{energy-density-prob}
			\end{equation}
			In plain English, it is precisely this that makes it quite clear that, in order for the right-hand side of this last equality to be identified, for instance, as the energy of a photon, it must be valid that
			\begin{equation}
				\int \boldsymbol{\varphi } ^{\ast } \left( \boldsymbol{k} , t \right) \cdot \boldsymbol{\varphi } \left( \boldsymbol{k} , t \right) \ d \boldsymbol{k} = 1 \ . \label{normalization}
			\end{equation}
			In this way, by taking into account the need for this normalization condition (\ref{normalization}), it is imperative to conclude that (\ref{foton-energy-nd}) can actually be recognized as the energy of a photon as long as it is valid that
			\begin{equation*}
				 n \left( k \right) = \sqrt{\frac{\hbar ck}{2 \left( 2 \pi \right) ^{3} \varepsilon _{0}}} \ \ \textnormal{or, equivalently,} \ \ n \left( \omega \right) = \sqrt{\frac{\hbar \omega }{2 \left( 2 \pi \right) ^{3} \varepsilon _{0}}} \ .
			\end{equation*} 
		
		\subsection{\label{subsec032}Some considerations on the linear momentum}
		
			Although the results obtained in the last Subsection seem quite satisfactory for the purpose of recognizing (\ref{maxwell-schrodinger}) as a Schr\"{o}dinger equation, we still need to evaluate other things to validate such recognition and, therefore, claiming that Maxwell's equations in free space already provide some support for the quantum description of electromagnetic radiation. And one of those other things that still needs to be evaluated is, for example, the electromagnetic linear momentum \cite{jackson}
			\begin{equation}
				\boldsymbol{p} = \varepsilon _{0} \int \boldsymbol{E} \left( \boldsymbol{r} , t \right) \times \boldsymbol{B} \left( \boldsymbol{r} , t \right) \ d \boldsymbol{r} \label{momento-p}
			\end{equation}
			that is associated with this electromagnetic radiation. After all, if these equations (\ref{maxwell}) already provide some support for the quantum description of electromagnetic radiation, it is expected that the development of (\ref{momento-p}) reveals to us that the relation (\ref{incerteza}) can actually be interpreted as the same Heisenberg's uncertainty principle (\ref{incerteza-heisenberg}).
			
			In fact, given all the calculations that we have developed so far, it is valid to say that making such an evaluation is not difficult since, with the help of (\ref{delta-function}), the development of (\ref{momento-p}) shows us that
			\begin{eqnarray}
				\boldsymbol{p} \negthickspace & = & \negthickspace \varepsilon _{0} \int \int \boldsymbol{\mathcal{E}} \left( \boldsymbol{k} , t \right) \times \boldsymbol{\mathcal{B}} \left( \boldsymbol{k^{\prime }} , t \right) \int e^{i \left( \boldsymbol{k + k^{\prime }} \right) \boldsymbol{r}} \ d \boldsymbol{r} \hspace*{0.04cm} d \boldsymbol{k} \hspace*{0.04cm} d \boldsymbol{k^{\prime }} \notag \\
				& = & \negthickspace \left( 2 \pi \right) ^{3} \varepsilon _{0} \int \boldsymbol{\mathcal{E}} \left( \boldsymbol{k} , t \right) \times \boldsymbol{\mathcal{B}} \left( - \boldsymbol{k} , t \right) \ d \boldsymbol{k} \ . \label{momento-p-aux-1}
			\end{eqnarray}
			Thus, by considering that
			\begin{eqnarray*}
				\boldsymbol{\mathcal{E}} \left( \boldsymbol{k} , t \right) \times \boldsymbol{\mathcal{B}} \left( - \boldsymbol{k} , t \right) \negthickspace & = & \negthickspace \boldsymbol{\mathcal{E}} \left( \boldsymbol{k} , t \right) \times \left[ - \frac{i}{\left( ck \right) ^{2}} \ \boldsymbol{k} \times \boldsymbol{\dot{\mathcal{E}}} \left( - \boldsymbol{k} , t \right) \right] \\
				& = & \negthickspace - \frac{i}{\left( ck \right) ^{2}} \left\{ \left[ \boldsymbol{\mathcal{E}} \left( \boldsymbol{k} , t \right) \cdot \boldsymbol{\dot{\mathcal{E}}} \left( - \boldsymbol{k} , t \right) \right] \boldsymbol{k} - \underbrace{ \left[ \boldsymbol{\mathcal{E}} \left( \boldsymbol{k} , t \right) \cdot \boldsymbol{k} \right] } _{=0} \dot{\boldsymbol{\mathcal{E}}} \left( - \boldsymbol{k} , t \right) \right\} \\
				& = & \negthickspace - \frac{i}{\left( ck \right) ^{2}} \ \boldsymbol{k} \left[ \boldsymbol{\mathcal{E}} \left( \boldsymbol{k} , t \right) \cdot \boldsymbol{\dot{\mathcal{E}}} \left( - \boldsymbol{k} , t \right) \right] \ ,
			\end{eqnarray*}
			and that the relations (\ref{fabulous-electric}), (\ref{e-derivadas-f}) and (\ref{f-first-order}), together, lead us to
			\begin{eqnarray}
				\boldsymbol{\mathcal{E}} \left( \boldsymbol{k} , t \right) \cdot \boldsymbol{\dot{\mathcal{E}}} \left( - \boldsymbol{k} , t \right) \negthickspace & = & \negthickspace - \frac{i}{2} \ \frac{\hbar \left( ck \right) ^{2}}{\left( 2 \pi \right) ^{3} \varepsilon _{0}} \ \bigl[ \boldsymbol{\varphi } \left( \boldsymbol{k} , t \right) \cdot \boldsymbol{\varphi } \left( - \boldsymbol{k} , t \right) - \boldsymbol{\varphi } \left( \boldsymbol{k} , t \right) \cdot \boldsymbol{\varphi } ^{\ast } \left( \boldsymbol{k} , t \right) \bigr] \notag \\
				& & \negthickspace - \frac{i}{2} \ \frac{\hbar \left( ck \right) ^{2}}{\left( 2 \pi \right) ^{3} \varepsilon _{0}} \ \bigl[ \boldsymbol{\varphi } ^{\ast } \left( - \boldsymbol{k} , t \right) \cdot \boldsymbol{\varphi } \left( - \boldsymbol{k} , t \right) - \boldsymbol{\varphi } ^{\ast } \left( - \boldsymbol{k} , t \right) \cdot \boldsymbol{\varphi } ^{\ast } \left( \boldsymbol{k} , t \right) \bigr] \ , \label{aux-mult} 
			\end{eqnarray}
			it is immediate to observe that (\ref{momento-p-aux-1}) reduces to
			\begin{eqnarray}
				\boldsymbol{p} \negthickspace & = & \negthickspace - \frac{\hbar }{2} \int \boldsymbol{k} \bigl[ \boldsymbol{\varphi } \left( \boldsymbol{k} , t \right) \cdot \boldsymbol{\varphi } \left( - \boldsymbol{k} , t \right) - \boldsymbol{\varphi } \left( \boldsymbol{k} , t \right) \cdot \boldsymbol{\varphi } ^{\ast } \left( \boldsymbol{k} , t \right) \bigr] \ d \boldsymbol{k} \notag \\
				& & \negthickspace - \frac{\hbar }{2} \int \boldsymbol{k} \bigl[ \boldsymbol{\varphi } ^{\ast } \left( - \boldsymbol{k} , t \right) \cdot \boldsymbol{\varphi } \left( - \boldsymbol{k} , t \right) - \boldsymbol{\varphi } ^{\ast } \left( - \boldsymbol{k} , t \right) \cdot \boldsymbol{\varphi } ^{\ast } \left( \boldsymbol{k} , t \right) \bigr] \ d \boldsymbol{k} \ , \label{momento-p-aux-2}
			\end{eqnarray}
			and, therefore, to
			\begin{equation}
				\boldsymbol{p} = \hbar \int \boldsymbol{k} \left[ \boldsymbol{\varphi } ^{\ast } \left( \boldsymbol{k} , t \right) \cdot \boldsymbol{\varphi } \left( \boldsymbol{k} , t \right) \right] \ d \boldsymbol{k} \label{momento-quantum}
			\end{equation}
			after a change $ \boldsymbol{k} \rightarrow - \boldsymbol{k} $ in the last integral appearing in (\ref{momento-p-aux-2}).
		
	\section{\label{sec04}What do these results tell us about the postulates of Quantum Mechanics, and vice versa?}
	
		Note that this result (\ref{momento-quantum}) is exactly what we needed to be able to interpret (\ref{incerteza}) as the Heisenberg's uncertainty principle (\ref{incerteza-heisenberg}) in this electrodynamic context. After all, as (\ref{wave-function}) shows us that
		\begin{equation}
			\int \boldsymbol{\varphi } ^{\ast } \left( \boldsymbol{k} , t \right) \cdot \boldsymbol{\varphi } \left( \boldsymbol{k} , t \right) \ d \boldsymbol{k} = \int \boldsymbol{\varphi } ^{\ast } _{0} \hspace*{-0.04cm} \left( \boldsymbol{k} \right) \cdot \boldsymbol{\varphi } _{0} \hspace*{-0.04cm} \left( \boldsymbol{k} \right) \ d \boldsymbol{k}
		\end{equation}
		and, therefore, that $ \boldsymbol{\varphi } \left( \boldsymbol{k} , t \right) $ and $ \boldsymbol{\varphi } _{0} \hspace*{-0.04cm} \left( \boldsymbol{k} \right) $ are square integrable functions \cite{rudin-1,rudin-2}, this result (\ref{momento-quantum}) indicates to us that it is possible to interpret $ \boldsymbol{k} \left[ \boldsymbol{\varphi } ^{\ast } \left( \boldsymbol{k} , t \right) \cdot \boldsymbol{\varphi } \left( \boldsymbol{k} , t \right) \right] $ as the probability density of finding a photon with a wave vector $ \boldsymbol{k} $. In other words, since the normalization condition (\ref{normalization}) further supports the interpretation of
		\begin{equation}
			\boldsymbol{\varphi } ^{\ast } \left( \boldsymbol{k} , t \right) \cdot \boldsymbol{\varphi } \left( \boldsymbol{k} , t \right) = \boldsymbol{\varphi } ^{\ast } _{0} \hspace*{-0.04cm} \left( \boldsymbol{k} \right) \cdot \boldsymbol{\varphi } _{0} \hspace*{-0.04cm} \left( \boldsymbol{k} \right) \label{densitity-prob-equal}
		\end{equation}
		as a probability density, it is precisely this result (\ref{momento-quantum}) that is showing us, for example, that $ \boldsymbol{p} $ can be interpreted as the expected value for the linear momentum of a photon. In this fashion, and by noting that this proportionality relationship (\ref{momento-quantum}) that exists between $ \boldsymbol{p} $ and $ \boldsymbol{k} $ leads us to
		\begin{equation}
			\Delta \boldsymbol{p} = \hbar \ \Delta \boldsymbol{k} \ , \label{p-hk}
		\end{equation}
		there is no way not to recognize that there is actually a correspondence, between the Maxwell's equations in free space and this new formulation (\ref{sistema-maxwell-schrodinger}), that is leading us to a quantum interpretation of these same equations. 
		
		Of course, a more attentive reader may wonder about the interpretation of the products in (\ref{densitity-prob-equal}) as a probability densities since, if this interpretation is indeed correct, it needs to appear in other results, and not just in (\ref{energy-density-prob}) and (\ref{momento-quantum}), because energy and momentum are not the only observables that exist. And, in order to clarify this important question, we invite this more attentive reader to evaluate, for example, the results that follow from manipulating the expression		
		\begin{equation}
			\boldsymbol{M} = \varepsilon _{0} \int \boldsymbol{r} \times \left[ \boldsymbol{E} \left( \boldsymbol{r} , t \right) \times \boldsymbol{B} \left( \boldsymbol{r} , t \right) \right] d \boldsymbol{r} \label{momento-angular}
		\end{equation}
		of the total angular momentum of this electromagnetic radiation described by ($ \boldsymbol{\varphi } \left( \boldsymbol{k} , t \right) $. After all, among the various things that this evaluation (which can be see, for instance, in Ref. \cite{akhiezer}) clearly shows us are the facts that:
		\begin{itemize}
			\item[$ \boldsymbol{\# 1} $] In addition to energy and linear momentum of electromagnetic radiation, all other physical observables $ A $, which can be identified with the manipulation of (\ref{momento-angular}), can also be associated with Hermitian operators $ \hat{A} $ such that
			\begin{equation}
				\left\langle A \hspace*{0.04cm} \right\rangle = \int \boldsymbol{\varphi } ^{\ast } \left( \boldsymbol{k} , t \right) \cdot \left[ \hat{A} \hspace*{0.04cm} \boldsymbol{\varphi } \left( \boldsymbol{k} , t \right) \right] \ d \boldsymbol{k} \ . \label{a-mean-integral}
			\end{equation}
			Here, $ \left\langle A \hspace*{0.04cm} \right\rangle $ denotes the expectation value of $ A $, at any time $ t $, as long as $ \boldsymbol{\varphi } \left( \boldsymbol{k} , t \right) $ is normalized.
			\item[$ \boldsymbol{\# 2} $] As with the energy and linear momentum of electromagnetic radiation, each of the other physical observables $ A $ related to item $ \boldsymbol{\# 1} $ has a definite value in $ \boldsymbol{\varphi } \left( \boldsymbol{k} , t \right) $ if, and only if, $ \boldsymbol{\varphi } \left( \boldsymbol{k} , t \right) $ is an eigenfunction of $ \hat{A} $. In other words, $ A $ has a definite value in $ \boldsymbol{\varphi } \left( \boldsymbol{k} , t \right) $ if, and only if,
			\begin{equation}
				\hat{A} \hspace*{0.04cm} \boldsymbol{\varphi } \left( \boldsymbol{k} , t \right) = a \hspace*{0.04cm} \boldsymbol{\varphi } \left( \boldsymbol{k} , t \right) \label{eigen-general-expression}
			\end{equation}
			because the only values that can arise, as a result of measuring $ A $, are the eigenvalues of $ \hat{A} $.
		\end{itemize}
		And since all other results, which can also be obtained by manipulating expressions other than (\ref{momento-angular}), allow us to recognise other physical observables that also satisfy $ \boldsymbol{\# 1} $ and $ \boldsymbol{\# 2} $, it is not wrong to conclude that the products contained in (\ref{densitity-prob-equal}) can, in fact, be interpreted as probability densities in this context, which is that of describing electromagnetic radiation in $ \boldsymbol{k} $-space \cite{akhiezer,mcweeny}.
		
		Incidentally, since $ \boldsymbol{\# 1} $ and $ \boldsymbol{\# 2} $ refer to these other physical observables that can be identified by manipulating (\ref{momento-angular}), a rather natural question arises: what are these observables? And, as simple as it may be to answer this question, doing so is very important because its answer leads us to an even more fundamental question, which is deeply connected to everything that has been presented so far. And what is this even more fundamental question? It is the one that deals with the spin that can be associated with electromagnetic radiation. After all, as much as Schrodinger obtained his wave equation by exploring a non-relativistic context \cite{schrodinger-1,schrodinger-2}, it is important to note that Maxwell's equations in free space, and consequently (\ref{sistema-maxwell-schrodinger}), are describing the behaviour of electromagnetic radiation that is relativistic. In this sense, and since the spin of a particle can always be interpreted as intrinsic angular momentum, it becomes quite intuitive to evaluate whether, for example, the manipulation of (\ref{momento-angular}) reveals anything about the spin of this electromagnetic radiation described by $ \boldsymbol{\varphi } \left( \boldsymbol{k} , t \right) $. And what this manipulation shows us is that (\ref{momento-angular}) can be rewritten in terms of
		\begin{equation}
			\boldsymbol{M} = \int \varphi ^{\ast } \left( \boldsymbol{k} , t \right) \cdot \bigl[ \boldsymbol{J} \hspace*{0.04cm} \varphi \left( \boldsymbol{k} , t \right) \bigr] \ d \boldsymbol{k} \ , \label{average-interpretation-ang-momentum}
		\end{equation}
		which only reinforces that $ \boldsymbol{M} $ really satisfies what was said in $ \boldsymbol{\# 1} $ and $ \boldsymbol{\# 2} $ because, for example, it can also be seen as the average value of the action of operator $ \boldsymbol{J} $ on the wave function $ \boldsymbol{\varphi } \left( \boldsymbol{k} , t \right) $.
		
		\subsection{Some considerations on the angular momentum and spin operators}
		
			Based on what we said in the last paragraph, the question that needs to be answered now is: what does this result (\ref{average-interpretation-ang-momentum}) have to do with the spin that can be associated with the electromagnetic radiation described by $ \boldsymbol{\varphi } \left( \boldsymbol{k} , t \right) $? And according to what is proven, for instance, in Ref. \cite{akhiezer}, the answer to this question is based on the fact that the manipulations that lead to (\ref{average-interpretation-ang-momentum}) make it very clear that
			\begin{equation}
				\boldsymbol{J} = \boldsymbol{L} + \boldsymbol{S} \ , \label{total-angular}
			\end{equation}
			where
			\begin{equation}
				\boldsymbol{L} = - i \hbar \ \left( \boldsymbol{k} \times \boldsymbol{\nabla } _{\boldsymbol{k}} \right) \label{orbital-m}
			\end{equation}
			has exactly the same expression of the orbital angular momentum of the Quantum Mechanics when expressed in $ \boldsymbol{k} $-space \cite{liboff}, and the action of $ \boldsymbol{S} $ on $ \boldsymbol{\varphi } \left( \boldsymbol{k} , t \right) $ can be, for example, represented with the help of matrices
			\begin{equation}
				\mathbb{S} _{1} = \begin{pmatrix}
					0 & 0 & 0 \\
					0 & 0 & -i \hbar \\
					0 & i \hbar & 0
				\end{pmatrix} \ , \quad \mathbb{S} _{2} = \begin{pmatrix}
					0 & 0 & i \hbar \\
					0 & 0 & 0 \\
					- i \hbar & 0 & 0
				\end{pmatrix} \quad \textnormal{and} \quad \mathbb{S} _{3} = \begin{pmatrix}
					0 & - i \hbar & 0 \\
					i \hbar & 0 & 0 \\
					0 & 0 & 0 
				\end{pmatrix} \label{s-matrices}
			\end{equation}
			as long as the matrix representation
			\begin{equation}
				\boldsymbol{\varphi } \left( \boldsymbol{k} , t \right) = \begin{pmatrix}
					\varphi _{1} \left( \boldsymbol{k} , t \right) \\
					\varphi _{2} \left( \boldsymbol{k} , t \right) \\
					\varphi _{3} \left( \boldsymbol{k} , t \right)
				\end{pmatrix} \label{onda-matrix}
			\end{equation}
			is also valid.
				
			Note that, as it is well known that operator $ \boldsymbol{L} = \left( \mathbb{L} _{1} , \mathbb{L} _{2} , \mathbb{L} _{3} \right) $ is, for instance, such that
			\begin{eqnarray}
				\boldsymbol{L} ^{2 \ } \hspace*{0.04cm} Y_{\ell m} \left( \boldsymbol{\theta } \right) & = & \ell \left( \ell + 1 \right) \hspace*{0.04cm} \hbar ^{2 \ } \hspace*{0.04cm} Y_{\ell m_{\ell }} \left( \boldsymbol{\theta } \right) \quad \textnormal{and} \notag \\
				\mathbb{L} _{3 \ } \hspace*{0.04cm} Y_{\ell m_{\ell }} \left( \boldsymbol{\theta } \right) & = & m_{\ell } \hspace*{0.04cm} \hbar \ Y_{\ell m_{\ell }} \left( \boldsymbol{\theta } \right) \ , \label{harmonics}
			\end{eqnarray}
			where $ Y_{\ell m_{\ell }} \left( \boldsymbol{\theta } \right) $ are spherical harmonic functions \cite{courant}, it is not wrong to state that the action of $ \boldsymbol{L} $ is restricted, here, only to the parameter $ \boldsymbol{\theta } = \left( \theta _{1} , \theta _{2} \right) $ whose unit vector is perpendicular to $ \boldsymbol{k} $ \footnote{Although it is usual to denote the spherical harmonic functions as $ Y_{\ell m_{\ell }} \left( \theta , \phi \right) $ (since the polar and azimuthal angles are popularly represented by the Greek letters $ \theta $ and $ \phi $ respectively), we preferred to write them as $ Y_{\ell m_{\ell }} \left( \theta _{1} , \theta _{2} \right) $ for the sake of a better understanding. After all, since we have already denoted an arbitrary infinitesimal rotation as $ \delta \boldsymbol{\theta } = \left( \delta \theta _{1} , \delta \theta _{2} \right) $, we needed to make it clear to you, the reader, that such a rotation is performed over the same subspace where such functions are defined.}. After all, when we notice that the components of $ \boldsymbol{k} $ take the form
			\begin{eqnarray}
				k_{1} & = & \kappa \left( \sin \theta _{1} \right) \left( \cos \theta _{2} \right) \ , \\
				k_{2} & = & \kappa \left( \sin \theta _{1} \right) \left( \sin \theta _{2} \right) \quad \textnormal{and} \\
				k_{3} & = & \kappa \left( \cos \theta _{1} \right)
			\end{eqnarray}
			when expressed by using spherical coordinates $ \left( \kappa , \boldsymbol{\theta } \right) $, it is precisely this that, when analysed from the perspective of the relations contained in (\ref{harmonics}), allows us to recognize, for example, that \cite{elon}
			\begin{equation}
				\varphi _{\alpha } \left( \boldsymbol{k} , t \right) = \sum ^{\infty } _{\ell = 0} \sum ^{m_{\ell } = \ell } _{m_{\ell } = - \ell } f_{\alpha \ell } \left( \kappa , t \right) Y _{\ell m_{\ell }} \left( \boldsymbol{\theta } \right) \label{desmonte}
			\end{equation}
			because the two indices $ \ell $ and $ m_{\ell } $, which appear in (\ref{harmonics}), are two integer numbers such that $ - \ell \leqslant m_{\ell } \leqslant \ell $. 
		
			Now, with regard to what has already been said about the action of $ \boldsymbol{S} = \left( \mathbb{S} _{1} , \mathbb{S} _{2} , \mathbb{S} _{3} \right) $ on $ \boldsymbol{\varphi } \left( \boldsymbol{k} , t \right) $, it is interesting to note that one of the useful consequences of dealing with this representation (\ref{s-matrices}) is that, with it, it is not difficult to see that all components $ \mathbb{S} _{\alpha } $ have three distinct eigenvalues, which are equal to $ - \hbar $, $ 0 $ and $ + \hbar $. As a consequence, another thing that is not difficult to notice is that, in the case of component $ \mathbb{S} _{3} $, its three normalized eigenstates, and respectively associated with these three eigenvalues, can be expressed as
			\begin{subequations} \label{three-h}
				\begin{align}
					\boldsymbol{s} _{- \hbar } \left( \boldsymbol{k} , t \right) & = \frac{1}{\sqrt{2}} \left[ \varphi _{1} \left( \boldsymbol{k} , t \right) \right] \boldsymbol{e} _{1} - \frac{i}{\sqrt{2}} \left[ \varphi _{2} \left( \boldsymbol{k} , t \right) \right] \boldsymbol{e} _{2} \ , \label{minus-h} \\
					\boldsymbol{s} _{0} \left( \boldsymbol{k} , t \right) & = \left[ \varphi _{1} \left( \boldsymbol{k} , t \right) \right] \boldsymbol{e} _{3} \quad \textnormal{and} \label{zero-h} \\
					\boldsymbol{s} _{+ \hbar } \left( \boldsymbol{k} , t \right) & = - \frac{1}{\sqrt{2}} \left[ \varphi _{1} \left( \boldsymbol{k} , t \right) \right] \boldsymbol{e} _{1} - \frac{i}{\sqrt{2}} \left[ \varphi _{2} \left( \boldsymbol{k} , t \right) \right] \boldsymbol{e} _{2} \ . \label{plus-h}
				\end{align}
			\end{subequations}
			Here,
			\begin{equation}
				\boldsymbol{e} _{1} = \begin{pmatrix}
					1 \\
					0 \\
					0
				\end{pmatrix} \ , \ \ \boldsymbol{e} _{2} = \begin{pmatrix}
					0 \\
					1 \\
					0
				\end{pmatrix} \ , \ \ \boldsymbol{e} _{3} = \begin{pmatrix}
					0 \\
					0 \\
					1
				\end{pmatrix} \label{base}
			\end{equation}
			are the matrix representations of the unit vectors that serve as the basis not only for (\ref{onda-matrix}), but also for, for example,
			\begin{equation}
				\boldsymbol{k} = \begin{pmatrix}
					k_{1} \\
					k_{2} \\
					k_{3}
				\end{pmatrix} \ . \label{k-representation}
			\end{equation}
			
			Because of not only the expressions of these eigenstates, but also the orthonormality in (\ref{base}), another thing that becomes clear here is that the two eigenstates (\ref{minus-h}) and (\ref{plus-h}) are actually describing two independent polarizations for a photon that is associated with electromagnetic radiation described by $ \boldsymbol{\varphi } \left( \boldsymbol{k} , t \right) $: more specifically, while (\ref{minus-h}) describes a left-handed circular polarization, (\ref{plus-h}) is describing a right-handed one. And when we also remember that, for such a wave function $ \boldsymbol{\varphi } \left( \boldsymbol{k} , t \right) $, the relation (\ref{f-perpendicular}) must also be valid, this is what explains why, for instance, nothing similar can be associated with the eigenstate (\ref{zero-h}). After all, as the relation (\ref{f-perpendicular}) only exists because $ \boldsymbol{k} $, by defining the direction of linear momentum, is always orthogonal to the components of the electromagnetic field that satisfy (\ref{remaxwell}), the matrix representation (\ref{k-representation}) makes it quite clear that
			\begin{itemize}
				\item the only components that describe such a field in this basis (\ref{base}) (and, therefore, lead to two independent circular polarizations) are $ \varphi _{1} \left( \boldsymbol{k} , t \right) $ and $ \varphi _{2} \left( \boldsymbol{k} , t \right) $, and
				\item the $ \varphi _{3} \left( \boldsymbol{k} , t \right) $ is not circularly polarizable in this basis (\ref{base}) because, in this basis, this component needs to be zero since, due to the condition (\ref{f-perpendicular}), it has the same direction as $ \boldsymbol{k} = \left( 0 , 0 , k_{3} \right) = \left( 0 , 0 , k \right) $.
			\end{itemize}
			In view of this last comment, it becomes valid to identify the unit vector $ \boldsymbol{e} _{3} $ as $ \boldsymbol{k} / k $. And one of the direct consequences of such identification is that, as it leads us to \cite{gomes}
			\begin{equation*}
				\boldsymbol{e} _{1, \alpha } \cdot \boldsymbol{e} _{1, \beta } + \boldsymbol{e} _{2, \alpha } \cdot \boldsymbol{e} _{2, \beta } = \delta _{\alpha \beta } - \frac{k_{\alpha} k_{\beta }}{k^{2}} \ , 
			\end{equation*}				
			the comparison of this new result with (\ref{maxwell-schrodinger-componentes}) makes clear that
			\begin{equation*}
				\mathbb{H} _{\alpha \beta } = \hbar ck \sum _{\lambda = 1,2} \boldsymbol{e} _{\lambda , \alpha } \cdot \boldsymbol{e} _{\lambda , \beta } \ .
			\end{equation*}
			In other words, this last result is showing us, in a very clear way, that the energy of a photon, whose movement is in the $ \boldsymbol{k} $ direction, is due solely and exclusively to the $ \varphi _{1} \left( \boldsymbol{k} , t \right) $ and $ \varphi _{2} \left( \boldsymbol{k} , t \right) $ components of the wave function that describes it.
		
		\subsection{Some considerations on some commutation relations}
				
			By the way, and by exploring this same context, it is also quite curious to note that the eigenvalues $ - \hbar $ and $ + \hbar $, which are respectively associated with (\ref{minus-h}) and (\ref{plus-h}), are exactly the same as those associated with the helicity (i.e., the projection of the spin) of a photon (on the direction of its linear momentum $ \boldsymbol{p} $) \cite{halzen}. And why is it quite curious to note this? Because it is precisely this that reinforces, even more, the consequent interpretation of $ \boldsymbol{S} $ as the operator that measures the spin of a photon since, for instance, these three eigenvalues of $ \mathbb{S} _{3} $ also satisfy a relation that is quite analogous to (\ref{harmonics}). After all, as the result
			\begin{equation*}
				\boldsymbol{S} ^{2} = \left( \mathbb{S} _{1} \right) ^{2} + \left( \mathbb{S} _{2} \right) ^{2} + \left( \mathbb{S} _{3} \right) ^{2} = \begin{pmatrix}
					2 \hbar ^{2} & 0 & 0 \\
					0 & 2 \hbar ^{2} & 0 \\
					0 & 0 & 2 \hbar ^{2}
				\end{pmatrix}
			\end{equation*}
			also makes clear that the three eigenvalues of $ \boldsymbol{S} ^{2} $ are all equal to $ 2 \hbar ^{2} $, it is not at all difficult to recognize that this complete analogy, between
			\begin{eqnarray}
				\boldsymbol{S} ^{2 \ } \hspace*{0.04cm} \boldsymbol{\varphi } \left( \boldsymbol{k} , t \right) & = & s \left( s + 1 \right) \hspace*{0.04cm} \hbar ^{2 \ } \hspace*{0.04cm} \boldsymbol{\varphi } \left( \boldsymbol{k} , t \right) \quad \textnormal{and} \notag \\
				\mathbb{S} _{3 \ } \hspace*{0.04cm} \boldsymbol{\varphi } \left( \boldsymbol{k} , t \right) & = & m_{s} \hspace*{0.04cm} \hbar \ \boldsymbol{\varphi } \left( \boldsymbol{k} , t \right) \ , \label{analogous-spin}
			\end{eqnarray}
			and all the results contained in (\ref{harmonics}), as long as it is valid that $ s = 1 $ since this guarantees that all possible integer values for $ m_{s} $ are actually such that $ - s \leqslant m_{s} \leqslant s $. In other words, even though we have made purely algebraic manipulations by taking, as a starting point, the expression for angular momentum (\ref{momento-angular}) that comes from Maxwell's electromagnetic theory, it is quite clear that we have obtained results that allow us to recognize (\ref{total-angular}) as the operator that measures the total angular momentum of a photon described by $ \boldsymbol{\varphi } \left( \boldsymbol{k} , t \right) $. And a very simple way to prove this is by noting that, since the operators $ \boldsymbol{L} $ and $ \boldsymbol{S} $ act on parameters of disjoint subspaces, these operators are necessarily such that
			\begin{equation*}
				\left[ \boldsymbol{L} ^{2} , \boldsymbol{S} ^{2} \right] = \left[ \mathbb{L} _{\alpha } , \boldsymbol{S} ^{2} \right] = \left[ \boldsymbol{L} ^{2} , \mathbb{S} _{\beta } \right] = \left[ \mathbb{L} _{\alpha } , \mathbb{S} _{\beta } \right] = 0 \ . 
			\end{equation*}
			After all, since the expressions in (\ref{orbital-m}) and (\ref{s-matrices}) allows to conclude that
			\begin{equation*}
				\left[ \mathbb{L} _{\alpha } , \mathbb{L} _{\beta } \right] = i \hbar \epsilon _{\alpha \beta \gamma } \mathbb{L} _{\gamma } \quad \textnormal{and} \quad \left[ \mathbb{S} _{\alpha } , \mathbb{S} _{\beta } \right] = i \hbar \epsilon _{\alpha \beta \gamma } \mathbb{S} _{\gamma } \ ,
			\end{equation*}
			it is precisely this that allows us to demonstrate the analogous result
			\begin{equation*}
				\left[ \mathbb{J} _{\alpha } , \mathbb{J} _{\beta } \right] = i \hbar \epsilon _{\alpha \beta \gamma } \mathbb{J} _{\gamma } 
			\end{equation*}
			that, for example, leads us to
			\begin{eqnarray}
				\boldsymbol{J} ^{2 \ } \hspace*{0.04cm} \boldsymbol{\varphi } \left( \boldsymbol{k} , t \right) & = & j \left( j + 1 \right) \hspace*{0.04cm} \hbar ^{2 \ } \hspace*{0.04cm} \boldsymbol{\varphi } \left( \boldsymbol{k} , t \right) \quad \textnormal{and} \notag \\
				\mathbb{J} _{3 \ } \hspace*{0.04cm} \boldsymbol{\varphi } \left( \boldsymbol{k} , t \right) & = & m_{j} \hspace*{0.04cm} \hbar \ \boldsymbol{\varphi } \left( \boldsymbol{k} , t \right) \ , \label{analogous-total-momentum}
			\end{eqnarray}
			where $ -j \leqslant m_{j} \leqslant j $.
			
			Given all the results presented so far, it is impossible not to recognise that, in addition to all the observables discussed here satisfying $ \boldsymbol{\# 1} $ and $ \boldsymbol{\# 2} $, the latest commutation relations allow us to state, for example, that:
			\begin{itemize}
				\item[$ \boldsymbol{\# 3} $] The operators $ \mathbb{H} $, $ \boldsymbol{L} ^{2 \ } $, $ \boldsymbol{S} ^{2 \ } $, $ \boldsymbol{J} ^{2 \ } $ and $ \mathbb{J} _{3 \ } $ defined above form, together, a set of Hermitian operators that commute two by two. That is, the commutation relation
				\begin{equation}
					\bigl[ \hat{A} , \hat{B} \bigr] = 0
				\end{equation}
				is always verified when $ \hat{A} $ and $ \hat{B} $ are identified with any of these operators \cite{zwiebach}.
			\end{itemize}
			In this fashion, as everything that was said in Section \ref{sec03} also allows us to assert that
			\begin{itemize}
				\item $ \boldsymbol{\varphi } \left( \boldsymbol{k} , t \right) $ belongs to an example of Hilbert space because it is square integrable \cite{rudin-1,rudin-2}, and 
				\item the state of electromagnetic radiation is completely determined by this same function $ \boldsymbol{\varphi } \left( \boldsymbol{k} , t \right) $, whose temporal evolution is governed by differential equation (\ref{maxwell-schrodinger}), where the operator $ \mathbb{H} $, which plays a leading role in this equation, can be recognised as a Hamiltonian operator,
			\end{itemize}
			it is also impossible not to recognise that Maxwell's electromagnetic theory already supports the quantum description of electromagnetic radiation \cite{akhiezer}. In other words, everything we have just observed here only proves that, even though Maxwell formulated his electromagnetic field theory without any quantum concerns, it is impossible not to recognize that all the results we have just presented are endorsing that this electromagnetic field theory in free space is defining an example of quantum theory. More specifically, all the results we have just presented, by analysing the electromagnetic field theory in free space, are not only defining the quantum mechanics of the photon, but also are pointing to the entire basis of a Quantum Mechanics that was only be formulated half a century later \cite{akhiezer,mcweeny}. Note that this is precisely what explains, for example, why many textbooks on quantum field theories always introduce such theories by manipulating Maxwell's equations in free space without giving in-depth explanations as to why they do so.

	\section{\label{sec05}Remarks on the foundations of quantum theories}
		
		Of course, due to the algebraic manipulations we made in items \textbf{(i)} and \textbf{(ii)}, a more attentive reader could counter-argue what we just said in the last paragraph. After all, since these algebraic manipulations depended on multiplications by the same real constant $ a $, which was later taken as $ \hbar $, it gives the impression that the quantum nature of Maxwell's equations in free space only becomes apparent due to a dirty trick. And, at first glance, we cannot even disagree with this counter-argument because, by performing these multiplications, it really does seem that we are performing a procedure analogous to what we do when, for example, we want to tune a radio receiver in to station that plays our favourite songs. In other words, it gives the impression that there is a very specific and artificial way of `tuning' the quantum mechanics of a photon through Maxwell's equations in free space and, as this way is by choosing the `radio station' $ a = \hbar $, this seems to be a big dirty trick because, at the time Maxwell formulated his electromagnetic field theory, no one had the slightest knowledge of what this constant $ \hbar $.
		
		However, as relevant as this counter-argument may be, it is also very important to note that, although $ \hbar $ has a very specific value in the International System of Units, this system of units is not the only one that exists. And one of the other systems of units that exist is, for example, the Lorentz-Heaviside system where, among other things, both the speed of light $ c $ and this same Planck constant $ \hbar $ are reduced to the same dimensionless unitary constant for theoretical purposes \cite{silsbee}. Therefore, even though it may seem that the multiplications, described in items \textbf{(i)} and \textbf{(ii)}, were made to force the `tuning' of this Quantum Mechanics underlying Maxwell's equations in free space, what this new observation reveals to us is that, in fact, such multiplications were not even necessary for this purpose. In plain English, all these mathematical developments are showing us that the constant $ \hbar $ should not be interpreted as a `tuning knob' that, as if by magic, transforms a classical wave into a quantum particle: this constant $ \hbar $ should be interpreted as a mere universal dimensional scaling factor that maps the amplitude of the continuous classical field (energy per unit volume) into the discrete energy quantum of single-mode field excitations (i.e., $ E = \hbar \omega $). After all, note that all the mathematical developments presented above, which led us to an equation whose form is, today, credited to Schr\"{o}dinger, were made using only resources that were already known at the beginning of the 19th Century.
		
		Given what we have just said, you, the reader, may be feeling a little uncomfortable. After all, how to justify, in a more fundamental way, the fact that we arrived, for instance, at this equation (\ref{maxwell-schrodinger}), which has the same form/content as the wave equation that Schr\"{o}dinger would get about twenty years after Einstein's quantum interpretation of light and other electromagnetic radiations? That is, why do Maxwell's equations in free space, when rewritten in the form (\ref{maxwell-schrodinger}), have features that are so familiar to a Quantum Mechanics that would only be formulated almost half a century after these equations? And the best answer that we can give to these questions can be summed up in a single term: correspondence principle. 
		
		\subsection{Some words on the correspondence principle}
		
			Although this principle became popularly known only due to the works of Niels Bohr (1885 -- 1962) in the advent of Quantum Mechanics \cite{bohr-1}, it is not wrong to say that it was always respected by all those who helped to build the most diverse physical theories. By the way, it is also not wrong to say that this principle only received its name, in 1920, because, contrary to what happened before the 17th Century, the language of Physics in the 20th Century was already extremely mathematical. After all, it is not wrong to assert that, because Quantum Mechanics was emerging within this extremely mathematical scenario, Paul Dirac (1902 -- 1984) realised that this `correspondence' must also necessarily be related to the mathematical correspondence
			\begin{equation}
				\mathcal{F} : U_{A} \rightarrow V_{B} \subset U_{B} \label{correspondence}
			\end{equation}
			that had to exist between two physical theories: in the case, between a theory $ A $, which was already well established due to several experiments and (theoretical) considerations (as, for example, Classical Mechanics was), and another $ B $ that was beginning to be proposed/constructed (as, for example, Quantum Mechanics was). By the way, due to the work not only of Bohr, but mainly of Dirac (who, for instance, improved the definition of this correspondence principle a little further, by modelling quantum mechanical systems in correspondence with the classical Hamiltonian systems) \cite{dirac}, it became clear, at that point, that this correspondence $ \mathcal{F} $ had to be established so that \cite{bolotin}
			\begin{itemize}
				\item its domain $ U_{A} $ contained all the functions that already describe a physical system in the context of the theory $ A $, and
				\item its codomain $ U_{B} $ contained all the new functions that describe this same physical system in the context of the theory $ B $.
			\end{itemize}
			
			It is worth noting that, since $ \mathcal{F} $ is a mathematical correspondence that relates two sets containing functions, some current research has been dedicated to investigating whether this correspondence principle can be, for example, interpreted from the point of view of category theory \cite{bolotin}. But what is most important to note here is that, as every new theory $ B $ always seeks to answer the questions that $ A $ does not yet provide, the image $ V_{B} $ does not necessarily identify with $ U_{B} $. However, as the new theory $ B $ cannot contradict the correct predictions of $ A $, this seems to suggest that this mathematical correspondence needs to be established so that these theories give the same results when some limits are considered \cite{costa}.
			
			Note that, in view of this apparent suggestion, the most basic and general definition we can give of this principle of correspondence ultimately explains, for example, why:
			\begin{itemize}
				\item the results of Special Relativity are reduced to those of non-relativistic Mechanics (Newtonian Physics) when the speed of light is taken as infinity \cite{weinberg}; and
				\item Statistical Mechanics is able to reproduce all the results of Thermodynamics when the number of particles in a system is supposed to be infinite \cite{salinas}.
			\end{itemize}
			However, as much as these two examples seem to show that these correspondences exist between two physical theories defined in different contexts, it is worth noting that this difference is not always necessary. In other words, whilst the correspondence principle formulated by both Bohr and Dirac explores an asymptotic physical limit, in which quantum predictions converge towards classical results in the regime of high quantum numbers (i.e., when) $ n $ ($ \rightarrow \infty $) or on infinitesimally small scales of action (i.e., when $ \hbar \rightarrow 0 $), the most general (and robust) definition of the correspondence principle shows that it is not always necessary to explore such a limit. And a good example where this is quite evident is in the three laws that Isaac Newton (1643 -- 1727) chose to base his Mechanics. After all, if Newton had not thought it reasonable that his Mechanics should respect this correspondence principle, he would never have, for instance, incorporated the Galileo Galilei's (1564 -- 1642) inertia principle as the first of these laws \cite{newton,letter-newton-hooke,letters-newton}. Therefore, it is valid to say that this correspondence, which predicates the name of this principle, may be identified as an inclusion map in some cases: i.e., it may be identified as a correspondence
			\begin{equation}
				\mathcal{F} : U_{A} \hookrightarrow V _{B} \ , \label{inclusion-map}
			\end{equation}
			where domain and image are exactly the same set.
			
			\subsubsection{\label{subsub-iso}On the identification of a structural isomorphism}
				
				Another example, where a correspondence map between two theories is not governed by asymptotic limits, is the one that justifies why the functions of the Hamiltonian Mechanics can be obtained by means of a Legendre transformation on the functions of the Lagrangian Mechanics \cite{lag-ham}. And as all concepts in Physics are constantly evolving \cite{popper,kuhn}, note that pointing out that this map (which can be recognised as an equivalence map when the mechanical systems under evaluation are not degenerate) also exists is entirely consistent, for example, with the perspective of Modern Philosophy of Science on the concept of the principle of correspondence. After all, observe that, as Heinz Post (1911 -- 1996) argues, the correspondence between theories is rarely restricted to mere asymptotic limits: in plain English, Post's generalised correspondence principle states that a successor theory $ B $ retains the well-confirmed mathematical structures and heuristics of its predecessor $ A $, such that the core algebraic and kinematic frameworks are preserved throughout the change of theory \cite{post}.
				
				Incidentally, also as noted by Alisa Bokulich (n. 19XY), the mathematical correspondence between two theories often manifests itself as a structural isomorphism between the state spaces and the symmetry groups of different physical theories, even when their underlying ontologies are completely different \cite{bokulich-1}. Observe that this is precisely the case with the mathematical correspondence between Lagrangian and Hamiltonian mechanics of non-degenerate systems because, in this case, this correspondence takes the form of a diffeomorphism (i.e., an example of an isomorphism) \cite{marsden}. But, although the last paragraph was devoted to this equivalence between Lagrangian and Hamiltonian Mechanics, everything we said in this last paragraph also applies, for instance, to the results we obtained in the previous sections. After all, as the development we carried out in (\ref{wave-equation-h-aux}) makes it quite clear that the function (\ref{wave-function}) describes a simple harmonic oscillator, this is where it becomes interesting to recall that QED is founded on the canonical quantisation of such functions \cite{mandl}.
				
				In order to understand this last point better, it is important to remember that, as the function (\ref{wave-function}) describes a simple harmonic oscillator, it is not unreasonable to assert that $ \boldsymbol{\varphi } \left( \boldsymbol{k} , t \right) $ and $ \boldsymbol{\varphi } ^{\ast } \left( \boldsymbol{k} , t \right) $ describe the same physical reality as the two real functions $ \boldsymbol{q} \left( \boldsymbol{k} , t \right) $ and $ \boldsymbol{p} \left( \boldsymbol{k} , t \right) $, provided that they satisfy the relation
				\begin{equation}
					\boldsymbol{\varphi } \left( \boldsymbol{k} , t \right) = \sqrt{\frac{1}{2 a \omega }} \left[ \hspace*{0.04cm} \omega \hspace*{0.04cm} \boldsymbol{q} \left( \boldsymbol{k} , t \right) + i \boldsymbol{p} \left( \boldsymbol{k} , t \right) \hspace*{0.04cm} \right] \ \Leftrightarrow \  \boldsymbol{\varphi } ^{\ast } \left( \boldsymbol{k} , t \right) = \sqrt{\frac{1}{2 a \omega }} \left[ \hspace*{0.04cm} \omega \hspace*{0.04cm} \boldsymbol{q} \left( \boldsymbol{k} , t \right) - i \boldsymbol{p} \left( \boldsymbol{k} , t \right) \hspace*{0.04cm} \right] \ , \label{transf-can-aa}
				\end{equation}
				where $ a $ is the same real constant we referred to in items \textbf{(i)} and \textbf{(ii)}. Observe that, as this ensures that $ \boldsymbol{q} \left( \boldsymbol{k} , t \right) $ and $ \boldsymbol{p} \left( \boldsymbol{k} , t \right) $ can be interpreted as the canonically conjugate functions, which are capable of describing a harmonic oscillator intrinsically because its Poisson brackets are such that \cite{lemos-book}
				\begin{equation}
					\begin{aligned}
						& \bigl\{ \boldsymbol{q} \left( \boldsymbol{k} , t \right) , \boldsymbol{q} \left( \boldsymbol{k^{\prime }} , t \right) \bigr\} = 0 \ , \\
						& \bigl\{ \boldsymbol{p} \left( \boldsymbol{k} , t \right) , \boldsymbol{p} \left( \boldsymbol{k^{\prime }} , t \right) \bigr\} = 0 \quad \textnormal{and} \\
						& \bigl\{ \boldsymbol{q} \left( \boldsymbol{k} , t \right) , \boldsymbol{p} \left( \boldsymbol{k^{\prime }} , t \right) \bigr\} = \delta \left( \boldsymbol{k} , \boldsymbol{k^{\prime }} \right) \ ,
					\end{aligned} \label{cano-qp}
				\end{equation}
				it is important to note that the properties of these same Poisson brackets also allow us to verify that
				\begin{subequations} \label{can-a-123}
					\begin{align}
						& \bigl\{ \boldsymbol{\varphi } \left( \boldsymbol{k} , t \right) , \boldsymbol{\varphi } \left( \boldsymbol{k^{\prime }} , t \right) \bigr\} = 0 \ , \label{can-a-1} \\
						& \bigl\{ \boldsymbol{\varphi } ^{\ast } \left( \boldsymbol{k} , t \right) , \boldsymbol{\varphi } ^{\ast } \left( \boldsymbol{k^{\prime }} , t \right) \bigr\} = 0 \quad \textnormal{and} \label{can-a-2} \\
						& \bigl\{ \boldsymbol{\varphi } \left( \boldsymbol{k} , t \right) , \boldsymbol{\varphi } ^{\ast } \left( \boldsymbol{k^{\prime }} , t \right) \bigr\} = \frac{1}{ia} \hspace*{0.04cm} \delta \left( \boldsymbol{k} , \boldsymbol{k^{\prime }} \right) \ . \label{can-a-3}
					\end{align}
				\end{subequations}
				After all, whilst the canonicity (\ref{cano-qp}) of the real functions $ \boldsymbol{q} \left( \boldsymbol{k} , t \right) $ and $ \boldsymbol{p} \left( \boldsymbol{k} , t \right) $ can be used to determine the corresponding self-adjoint operators
				\begin{subequations}
					\begin{align*}
						\boldsymbol{q} \left( \boldsymbol{k} , t \right) & \ \mapsto \ \hat{Q} _{\boldsymbol{k}} \left( t \right) = \mathcal{Q} \left( \boldsymbol{q} \left( \boldsymbol{k} , t \right) \right) \\
						\boldsymbol{p} \left( \boldsymbol{k} , t \right) & \ \mapsto \ \hat{P} _{\boldsymbol{k}} \left( t \right) = \mathcal{Q} \left( \boldsymbol{p} \left( \boldsymbol{k} , t \right) \right)
					\end{align*}
				\end{subequations}
				in quantum theory \cite{dirac}, where $ \mathcal{Q} $ is an injective mapping that must satisfy the commutation relations
				\begin{equation}
					\lim _{\hbar \rightarrow 0} \left( \frac{1}{i \hbar } \left[ \hat{F} , \hat{G} \right] \right) = \bigl\{ F , G \bigr\} \label{quantum-condition}
				\end{equation}
				for any two operators $ \hat{F} = \mathcal{Q} \left( F \right) $ and $ \hat{G} = \mathcal{Q} \left( G \right) $ \cite{assirati}, transformation (\ref{transf-can-aa}) already makes it quite clear that relations (\ref{can-a-123}) also allow us to recognise
				\begin{subequations} \label{map-iso-nd}
					\begin{align}
						\boldsymbol{\varphi } \left( \boldsymbol{k} , t \right) & \ \mapsto \ \hat{a} _{\boldsymbol{k}} \left( t \right) = \mathcal{Q} \left( \boldsymbol{\varphi } \left( \boldsymbol{k} , t \right) \right) \label{map-iso-n} \\
						\boldsymbol{\varphi } ^{\ast } \left( \boldsymbol{k} , t \right) & \ \mapsto \ \hat{a} ^{\dagger } _{\boldsymbol{k}} \left( t \right) = \mathcal{Q} \left( \boldsymbol{\varphi } ^{\ast }  \left( \boldsymbol{k} , t \right) \right) \label{map-iso-d}
					\end{align}
				\end{subequations}
				as the quantum operators of this theory\footnote{Here, the operators $ \hat{a} _{\boldsymbol{k}} \left( t \right) $ and $ \hat{a} ^{\dagger } _{\boldsymbol{k}} \left( t \right) $ should be interpreted as the Hermitian conjugates (or adjoints) of one another \cite{mandl}.}. And when we realise that, by setting $ a = \hbar $, the denominator on the left-hand side of (\ref{can-a-3}) is the same as the one on the right-hand side of (\ref{quantum-condition}), here it becomes clear that the correspondence map, between Maxwell's field equations in free space and the wave equation (\ref{maxwell-schrodinger}), does not define a situation in which a Bohrian asymptotic limit can be observed. 
				
			\subsubsection{An important remark on this canonical quantisation}
			
				Of course, there are a few caveats that need to be noted regarding this isomorphism map (\ref{map-iso-nd}). One of them requires us to observe that, whilst Schrödinger's Quantum Mechanics holds that it is always possible to describe the physics of any system by means of a wave function written in $ \boldsymbol{r} $-space \cite{schrodinger-papers}, the Schrödinger equation (\ref{maxwell-schrodinger}) is restricted to $ \boldsymbol{k} $-space. After all, unlike Schrödinger's Quantum Mechanics, which is restricted to the description of non-relativistic massive particles, the Newton-Wigner theorem establishes that massless relativistic particles with non-zero spin (such as photons) do not possess a strictly local, commuting spatial position operator $ \boldsymbol{\hat{r}} $ in configuration space \cite{newton-wigner}. In this fashion, while $\boldsymbol{\varphi } \left( \boldsymbol{k} , t \right) $ serves as a rigorous, localized photon wave function in $ \boldsymbol{k} $-space, it is correct to say that its spatial counterpart $ \boldsymbol{\psi } \left( \boldsymbol{r} , t \right) $ cannot be interpreted as a point-like probability amplitude of photon localization without violating relativistic causality over sub-wavelength scales \cite{birula-1,birula-2,birula-3}.
				
				But what is particularly interesting to note here is that, once again, the same need to set $ a = \hbar $ resurfaces. After all, this resurgence further reinforces that QED is defined by implementing a mathematical correspondence (\ref{map-iso-nd}) that, for example, alters neither the oscillation mode space nor even the geometry of Maxwell's equations, since it simply maps the complex amplitudes $ \boldsymbol{\varphi } \left( \boldsymbol{k} , t \right) $ and $ \boldsymbol{\varphi } ^{\ast } \left( \boldsymbol{k} , t \right) $ of these classical harmonic oscillators onto the respective quantum operators $ \hat{a} _{\boldsymbol{k}} \left( t \right) $ and $ \hat{a} ^{\dagger } _{\boldsymbol{k}} \left( t \right) $ acting in Fock space \cite{fock}. That is to say, neither Classical Electrodynamics `becomes' Quantum Mechanics when the number of photons is small, nor does Quantum Mechanics `contain' Classical Electrodynamics as a mere subset. Instead, what we observe is an exact kinematic isomorphism between the classical Maxwell field equations in $ \boldsymbol{k} $-space and the single-photon sector of QED in Fock space.
				
			\subsubsection{Further remarks on this structural isomorphism}
				
				Observe that, in Post's sense, this Maxwell-Schrödinger correspondence is a case where the successor theory (i.e., quantum mechanics of a photon) retains the full algebraic and kinematic structure of the predecessor (i.e., Maxwell's electrodynamics in $ \boldsymbol{k} $-space), while adding a probabilistic interpretation and the quantum of action. In other words, and in line with everything we have observed in the previous Sections, Maxwell's equations in free space describe the temporal evolution of a classical vector field, whose underlying symmetry group (i.e., the Lorentz-invariant little group $ E \left( 2 \right) $ for massless particles \cite{wigner}), matches the state space of the quantum mechanics of a single photon with helicity $ \pm 1 $.
				
				One way for us to reinforce what we have just stated in the last paragraph is to recall what we have already said, for example, in facts $ \boldsymbol{\# 1} $, $ \boldsymbol{\# 2} $ and $ \boldsymbol{\# 3} $ announced in Section \ref{sec04}. After all, if we turn our attention, for instance, to what we have already said in $ \boldsymbol{\# 1} $, it is not incorrect to assert that both the normalisation condition (\ref{normalization}) and (consequently) the expression (\ref{a-mean-integral}) are the main ones responsible for endowing the space of transverse classical field modes with a standard inner product in the Hilbert space of square-integrable functions on the $ \boldsymbol{k} $-space \cite{rudin-1}. And, from the perspective of that exact kinematic isomorphism that we identified in Subsubsection \ref{subsub-iso}, this is precisely what, ultimately, skirts around the problem of the photon's non-locality in the configuration space, as established by the Newton-Wigner theorem.
				
				Now, with regard to what we said in $ \boldsymbol{\# 2} $, it is worth noting that relation (\ref{eigen-general-expression}), when analysed using the condition of transversality (\ref{f-perpendicular}), already makes it quite clear that the longitudinal state (\ref{zero-h}) can be classified as non-physical. Note that this is fully consistent with, for instance,
				\begin{itemize}
					\item the interpretation of Maxwell's electromagnetic theory as a gauge theory from the perspective of theories of constrained Hamiltonian systems \cite{gitman,mf-constraints}, and
					\item the fact that, from the perspective of irreducible unitary representations of the group $ E \left( 2 \right) $, the space of physical states of the electrodynamic system that we analysed in Sections \ref{sec03} and \ref{sec04} is, in fact, restricted only to the two eigenstates (\ref{minus-h}) and (\ref{plus-h}) \cite{wigner,bargmann}.
				\end{itemize}
				In this fashion, when we also take into account the fact that, in $ \boldsymbol{\# 3} $, we announced the existence of a complete set of Hermitian observables $ \left\{ \mathbb{H} , \boldsymbol{L} ^{2 \ } , \boldsymbol{S} ^{2 \ } , \boldsymbol {J} ^{2 \ } , \mathbb{J} _{3 \ } \right\} $, this is precisely what ultimately confirms that all the algebraic symmetries of Maxwell's  electrodynamics in the $ \boldsymbol{k} $-space, when there are no electric charges or currents, are, actually, fully capable of encoding the entire kinematic structure of the quantum mechanics of a single photon.
	
	\section{Final remarks}
	
		As important as it is to make further algebraic developments in order to reveal the quantum aspects that underlie Maxwell's equations where $ \left( c \rho \left( \boldsymbol{r} , t \right) , \boldsymbol{J} \left( \boldsymbol{r} , t \right) \right) \neq \left( 0 , \boldsymbol{0} \right) $, we will stop here and delegate such discussion to a future paper. After all, besides not wanting to make this paper even longer than it already is, what we have presented above makes it quite clear how Maxwell's electromagnetic theory in free space and the correspondence principle already pointed, together, to the basis of a Quantum Mechanics that was only formulated more than half a century after Maxwell's work.
		
		However, it is interesting to note that, according to the differential equations (\ref{dif-eq}), there already exists a situation that does not seem to be difficult to analyse: the one where the four-current density is defined by $ \rho \left( \boldsymbol{r} , t \right) = 0 $, but where $ \boldsymbol{J} \left( \boldsymbol{r} , t \right) $ satisfies the ohmic condition (\ref{ohmic}). And in order to understand why this is, it is enough to see that, in addition to equations (\ref{remaxwell-1}) and (\ref{remaxwell-2}) continuing to guarantee that $ \boldsymbol{\mathcal{E}} \left( \boldsymbol{k} , t \right) $, $ \boldsymbol{\mathcal{B}} \left( \boldsymbol{k} , t \right) $ and $ \boldsymbol{k} $ maintain the same perpendicularity relations as in free space, it also continues to be perfectly possible to deal with an expression
		\begin{equation*}
			\boldsymbol{\mathcal{E}} \left( \boldsymbol{k} , t \right) = n \left( k \right) \bigl[ \boldsymbol{\varphi } \left( \boldsymbol{k} , t \right) + \boldsymbol{\varphi } ^{\ast } \left( - \boldsymbol{k} , t \right) \bigr]
		\end{equation*}
		that, when substituted into (\ref{dif-eq-e}), makes it very clear that
		\begin{equation}
			\left[ \frac{\partial ^{2}}{\partial t^{2}} + \gamma \frac{\partial }{\partial t} + \omega ^{2} \right] \boldsymbol{\varphi } \left( \boldsymbol{k} , t \right) = 0 \ . \label{final-dif-equation}
		\end{equation}
		Regardless of what the other results in Section \ref{sec02} can show us about the expression that defines $ \boldsymbol{\mathcal{B}} \left( \boldsymbol{k} , t \right) $, it is interesting to highlight what follows, for example, from the use of slowly varying envelope approximation \cite{born-3} on this last equation. After all, when we analyse a situation where the function that solves (\ref{final-dif-equation}) can be expressed as
		\begin{equation*}
			\boldsymbol{\varphi } \left( \boldsymbol{k} , t \right) = \boldsymbol{\Upsilon } \left( \boldsymbol{k} , t \right) e^{- i \overline{\omega } t} \ , 
		\end{equation*}
		with the help of an envelope $ \boldsymbol{\Upsilon } \left( \boldsymbol{k} , t \right) $ whose average angular frequency $ \overline{\omega } $ that characterizes it is such that
		\begin{equation*}
			\left\vert \frac{\partial ^{2} \boldsymbol{\Upsilon }}{\partial t^{2}} \right\vert \ll \overline{\omega } \left\vert \frac{\partial \boldsymbol{\Upsilon }}{\partial t} \right\vert \ ,
		\end{equation*}
		it is not difficult to conclude that this leads us to
		\begin{equation*}
			\left( - 2i \overline{\omega } + \gamma \right) \frac{\partial \boldsymbol{\Upsilon }}{\partial t} + \left[ \left( \omega ^{2} - \overline{\omega } ^{2} \right) - i \gamma \overline{\omega } \right] \Upsilon \left( \boldsymbol{k} , t \right) \approx \boldsymbol{0} 
		\end{equation*}
		and, therefore, to
		\begin{equation}
			\frac{1}{2} \left( \frac{\omega ^{2} - \overline{\omega } ^{2}}{\overline{\omega }} - i \gamma \right) \Upsilon \left( \boldsymbol{k} , t \right) \approx i \frac{\partial \boldsymbol{\Upsilon }}{\partial t} \label{new-gamma-schrodinger}
		\end{equation} 
		in the special cases where $ \gamma \ll \overline{\omega } $. In this fashion, and even though we have already said that we will only analyse the quantum aspects that underlie Maxwell's equations where $ \left( c \rho \left( \boldsymbol{r} , t \right) , \boldsymbol{J} \left( \boldsymbol{r} , t \right) \right) \neq \left( 0 , \boldsymbol{0} \right) $ in a future paper, there is no way we can close this present paper without recognizing that, when we multiply both sides of equation (\ref{new-gamma-schrodinger}) by the same constant $ \hbar $, this leads us to an equation
		\begin{equation*}
			\frac{\hbar }{2} \left( \frac{\omega ^{2} - \overline{\omega } ^{2}}{\overline{\omega }} \right) \Upsilon \left( \boldsymbol{k} , t \right) - \frac{i \hbar \gamma }{2} \Upsilon \left( \boldsymbol{k} , t \right) \approx i \hbar \frac{\partial \boldsymbol{\Upsilon }}{\partial t} 
		\end{equation*}
		that, again, is very similar to Schr\"{o}dinger's, but which contains a damping factor. 
		
	\section{Acknowledgements}
	
		We thank J. L. M. Assirati for some physical and mathematical discussions on subjects concerning this paper. However, a very special thanks goes to D. M. Gitman who, after observing, in a conversation with M. F., that the quantum aspects underlying Maxwell's electrodynamics go unnoticed by some people, suggested that a detailed presentation of this fact would make an excellent pedagogical review paper. In fact, there are several references that can be used as a guide to better understand the quantum nature underlying this theory. And, among them, the one we find most interesting at the moment is Ref. \cite{akhiezer}, which we actually used as a guide to write some parts of this paper, especially its Sections \ref{sec03} and \ref{sec04}.

	\bigskip{\small \smallskip\noindent Updated: \today.}

\end{document}